\documentclass[11pt]{article}

\usepackage[latin1]{inputenc}
\usepackage{graphicx}
\usepackage{amsmath}
\usepackage{color}
\usepackage{graphicx}
\usepackage{hyperref}
\usepackage{comment}
\usepackage[mathscr]{euscript}
\usepackage{algorithm}
\usepackage[noend]{algpseudocode}
\usepackage{natbib}
\usepackage{amssymb}
\usepackage{multirow}
\usepackage{enumitem}
\usepackage{lscape}

\newcommand{\xxi}{\mbox{\boldmath{$\xi$}}}

\newcommand{\ttheta}{\mbox{\boldmath{$\theta$}}}

\newcommand{\ggamma}{\mbox{\boldmath{$\gamma$}}}
\newcommand{\ddelta}{\mbox{\boldmath{$\delta$}}}

\newcommand{\0}{\bf 0}

\newcommand{\rr}{\bf r}
\newcommand{\uu}{\bf u}
\newcommand{\vv}{\bf v}

\newcommand{\D}{\bf D}
\newcommand{\I}{\bf I}
\newcommand{\Q}{\bf Q}
\newcommand{\R}{\bf R}
\newcommand{\W}{\bf W}

\begin{document}


\title{\textbf{Big problems in spatio-temporal disease mapping: methods and software}}

\author{Orozco-Acosta, E$^{1,2}$, Adin, A$^{1,2}$ and Ugarte, M.D.$^{1,2}$\\
\small {\textit{$^1$ Department of Statistics, Computer Sciences and Mathematics, Public University of Navarre, Spain.}} \\
\small {\textit{$^2$ Institute for Advanced Materials and Mathematics, InaMat$^2$, Public University of Navarre, Spain.}}\\
\small {$*$Correspondence to María Dolores Ugarte, Departamento de Estad\'istica, Inform\'atica y Matem\'aticas, } \\
\small {Universidad P\'ublica de Navarra, Campus de Arrosadia, 31006 Pamplona, Spain.} \\
\small {\textbf{E-mail}: lola@unavarra.es }}
\date{}

\makeatletter
\pdfbookmark[0]{\@title}{title}
\makeatother

\maketitle

\begin{abstract}
Fitting spatio-temporal models for areal data is crucial in many fields such as cancer epidemiology. However, when data sets are very large, many issues arise. The main objective of this paper is to propose a general procedure to analyze high-dimensional spatio-temporal count data, with special emphasis on mortality/incidence relative risk estimation. We present a pragmatic and simple idea that permits to fit hierarchical spatio-temporal models when the number of small areas is very large. Model fitting is carried out using integrated nested Laplace approximations over a partition of the spatial domain. We also use parallel and distributed strategies to speed up computations in a setting where Bayesian model fitting is generally prohibitively time-consuming and even unfeasible. Using simulated and real data, we show that our method outperforms classical global models. We implement the methods and algorithms that we develop in the open-source \texttt{R} package \texttt{bigDM} where specific vignettes have been included to facilitate the use of the methodology for non-expert users. Our scalable methodology proposal provides reliable risk estimates when fitting Bayesian hierarchical spatio-temporal models for high-dimensional data.
\end{abstract}


\bigskip

\section{Introduction}
\label{intro}
In recent decades, access to geospatial data through Geographical Information Systems (GIS) and other related technologies has grown at a staggering rate. Modern geospatial data typically involve large datasets collected from a variety of sources (databases or servers) that may include information such as satellite imagery, weather data, census data, social network data and public health data.  Consequently, the development of new techniques and computational algorithms to analyze massive spatial and spatio-temporal datasets is of crucial interest in many fields such as remote sensing, geoscience, ecology, crime research and epidemiology among others.

Hierarchical spatial models including random effects \citep{cressie2011statistics,banerjee2015hierarchical} are widely used in spatial statistics to provide reliable estimates of the underlying geographical phenomenon and quantifying uncertainty in predictions at unobserved locations. See, for example, \cite{sun2012geostatistics} and \cite{banerjee2017high} for a detailed review of methods and scalable models for high-dimensional spatial and spatio-temporal data. Gaussian processes (GPs) have been commonly used for the analysis of geostatistical (point-referenced) data in the spatial statistics literature. However, traditional estimation of GPs has become computationally intractable when analysing modern big datasets, mainly due to computations involving matrix factorizations for very large covariance matrices. During the last years, many approaches have been proposed to ensure scalability of large geostatistical datasets (see, e.g., \cite{heaton2019case} and \cite{liu2020gaussian} for recent reviews and comparisons). Some other recent methods to deal with massive datasets are described below. \cite{appel2020spatiotemporal} provide an extension to the multi-resolution approximation approach \citep{katzfuss2017multi} for spatio-temporal modelling of global datasets, where a recursive partitioning scheme is considered so that inference can be efficiently scaled in distributed computing environments. \cite{zammit2020multi} propose an approximate inference scalable algorithm for multi-scale process modelling by using the stochastic partial differential equation approach \citep{lindgren2011explicit}. 
Both methods were applied to modelling and prediction of global sea-surface temperature.
In the geostatistical literature, many recent works are being proposed to estimate GPs based on the so-called Vecchia approximation. \citep{vecchia1988estimation}. This approximation can be regarded as a special case of the Gaussian Markov random field approximations  \citep{rue2005gaussian} with a simplified neighbourhood structure that can be represented by directed acyclic graph (DAG) models. This representation leads to a sparse formulation of the precision matrix which ensures that evaluating the likelihood of the GPs will be computationally scalable. Based on this approach, \cite{finley2019efficient} propose alternative formulations of Bayesian nearest neighbour Gaussian process models developed by \cite{datta2016hierarchical} to substantially improve computational efficiency; \cite{peruzzi2020highly} develop a meshed Gaussian process with a novel partitioning and graph design based on domain tessellations while \cite{katzfuss2021general} propose a novel sparse general Vecchia approximation algorithm which ensures computational feasibility for large spatial datasets; \cite{jurek2022} present a fast and simple algorithm to compute their hierarchical Vecchia approximation, and provide extensions to nonlinear data assimilation with non-Gaussian data based on the Laplace approximation.

Although there is an extensive literature developing scalable methods and computational algorithms for analysing massive geostatistical data, only a few papers discuss scalable disease mapping models for high dimensional areal data. Disease mapping is the field of spatial epidemiology that deals with aggregated count data from non-overlapping areal units focussing on the estimation of the geographical distribution of a disease and its evolution in time \citep{lawson2016handbook}.
As outlined by \cite{shen2000triple}, the three main inferential goals in disease mapping are: (i) to provide accurate estimates of mortality/incidence risks or rates in space and time, (ii) to unveil the underlying spatial and spatio-temporal patterns, and (iii) to detect high-risk areas or hotspots.
Since classical risk estimation measures, such as the standardized mortality/incidence ratio, are extremely variable when analysing rare diseases (with very few cases) or low-populated areas, several statistical models have been proposed during the last decades to obtain smooth disease risk estimates borrowing information from spatial and/or temporal neighbours. Research into spatial and spatio-temporal disease mapping has been carried out within a hierarchical Bayesian framework, with generalized linear mixed models (GLMMs) playing a major role. Although GLMMs including spatial and temporal random effects are a very popular and flexible approach to model areal count data, these smoothing methods become computationally challenging (or even unfeasible) when analysing very large spatio-temporal datasets.
\cite{guan2018computationally} develop a method to reduce the dimension of the spatial random effect by reparameterizing the model based on random projections of the covariance matrix. In addition, they show how to address confounding issues if explanatory variables are included in the model by simultaneously applying the restricted spatial regression approach \citep{hodges2010adding}. This model is similar to the one proposed by \cite{hughes2013dimension}, where the decomposition is performed based on the Moran operator.
\cite{datta2019spatial} introduce a class of directed acyclic graphical autoregression (DAGAR) models as an alternative to the commonly used conditional autoregressive (CAR) models for spatial areal data. Instead of modelling the precision matrix of the spatial random effect, they propose to model its (sparse) Cholesky factor using autoregressive covariance models on a sequence of local trees created from the directed acyclic graph derived from the original undirected graph (spatial neighbourhood structure) of the areal units. As stated by the authors, the Cholesky factor has the same level of sparsity as the undirected graph ensuring scalability for analysing very large areal datasets. An extension to deal with multivariate spatial disease mapping models has been developed by \cite{gao2021hierarchical}.
Very recently, a scalable Bayesian spatial model has been proposed by \cite{orozco2021scalable} based on the ``divide-and-conquer'' approach so that local spatial CAR models can be simultaneously fitted. This new methodology provides reliable risk estimates with a substantial reduction in computational time.

The modelling approaches described above are limited to the analysis of spatial count data. The main objective of this paper is to propose a scalable Bayesian modelling approach to smooth mortality or incidence risks in in a high-dimensional spatio-temporal disease mapping context by extending the methodology described in \cite{orozco2021scalable}.
Specifically, we adapt the modelling scheme so that commonly used spatio-temporal models can be fitted over different subdomains (partitions of the region of interest), which allows to define non-stationary models, i.e., models that induce different degree of smoothing over the areal units belonging to each subdomain. From a theoretical point of view, both spatial and/or temporal partitions of the data could be defined, however, in the disease mapping context the high-dimensionality of the data is usually related to the estimation of relative risks at a fine-scale spatial resolution. The main challenges of the methodology presented in this work is not only to extend the ``divide-and-conquer'' approach to deal with spatio-temporal models (which is not trivial at all), but also to derive and implement specific algorithms to perform scalable model estimation in both parallel or distributed processing architectures.

The remainder of this article is organized as follows. Section~\ref{sec:ST_models} poses the spatio-temporal CAR models considered in this work. Section~\ref{sec:Scalable_models} introduces the new scalable Bayesian models and describes a generic scheme of the main algorithms that have been implemented in this work. In Section~\ref{sec:SimulationStudy}, we conduct a simulation study based on a template of over the almost 8000 municipalities of continental Spain and 25 time periods to compare the new scalable methods with previous proposals.  In addition, we provide a numerical simulation  to evaluate the computational gain offered by our modelling approach when the number of small areas increases. In Section~\ref{sec:Data_analysis} we use the new model proposal to analyze lung cancer mortality data in Spanish municipalities. The paper ends with a discussion.

\section{Background: spatio temporal models in disease mapping}
\label{sec:ST_models}

Let us assume that the region under study is divided into contiguous small areas labelled as $i=1,\ldots,n$ and data are available for consecutive time periods labelled as $t=1,\ldots,T$. For a given area $i$ and time period $t$, $O_{it}$ and $E_{it}$ denote the number of observed and expected cases, respectively. To compute the number of expected cases both direct and indirect standardization methods can be used, usually considering age and/or sex as standardization variables. When using the indirect method, the number of expected cases for area $i$ and time $t$ is calculated as
\begin{equation*}
E_{it}=\sum\limits_{j=1}^J N_{it}\frac{O_j}{N_j} \quad \mbox{for} \; i=1,\ldots,n; \; t=1,\ldots,T,
\end{equation*}
where $O_j=\sum\limits_{i=1}^n\sum\limits_{t=1}^T O_{itj}$ and $N_j=\sum\limits_{i=1}^n\sum\limits_{t=1}^T N_{itj}$ are the number of observed cases and the population at risk in the $j^{th}$ age-and-sex group, respectively. Then, the standardized mortality/incidence ratio (SMR or SIR) is defined as the number of observed cases divided by the number of expected cases. Although its interpretation is very simple, SMRs are extremely variable when analysing rare diseases or very low-populated areas, as it is the case of high-dimensional data. This makes it necessary the use of statistical models to smooth risks borrowing information from neighbouring regions and time periods.

Poisson mixed models are typically used for the analysis of count data within a hierarchical Bayesian framework. Conditional to the relative risk $r_{it}$, the number of observed cases in the $i^{th}$ area and time period $t$ is assumed to be Poisson distributed with mean $\mu_{it}=E_{it}r_{it}$. That is,
\begin{eqnarray*}
\begin{array}{rcl}
O_{it} | r_{it} & \sim & Poisson(\mu_{it}=E_{it}r_{it}), \\
\log{\mu_{it}} & = & \log{E_{it}} + \log{r_{it}},\\
\end{array}
\end{eqnarray*}
where $\log{E_{it}}$ is an offset. Depending on the specification of the log-risks different models can be defined.

Probably, the non-parametric models based on CAR priors for spatial random effects, random walk priors for temporal random effects, and different types of spatio-temporal interactions described in \cite{KnorrHeld2000} are the most widely used models in space-time disease mapping. Slight modifications of these models are considered here, so the log-risks are modelled as
\begin{equation}
\label{eq:Global_model}
\log{r_{it}} = \alpha + \xi_i + \gamma_t + \delta_{it},
\end{equation}
where $\alpha$ is an intercept representing the overall log-risk, $\xi_i$ is a spatial random effect with CAR prior distribution, $\gamma_t$ is a temporally structured random effect that follows a random walk prior distribution, and $\delta_{it}$ is a spatio-temporal random effect. All the components of this model can be modelled as GMRFs and prior densities can be written according to some structure matrices.

A modification of the \cite{dean2001} model proposed by \cite{riebler2016intuitive}, hereafter called BYM2 model, has been considered as the prior distribution for the spatial random effects ${\xxi}=(\xi_1,\ldots,\xi_n)^{'}$, so that
\begin{equation*}
{\xxi}=\frac{1}{\sqrt{\tau_{\xi}}} \left(\sqrt{1-\lambda_{\xi}}{\vv} + \sqrt{\lambda_{\xi}}{{\uu}_{\star}} \right),
\end{equation*}
where $\tau_{\xi}$ is a precision parameter, $\lambda_{\xi} \in [0,1]$ is a spatial smoothing parameter, ${\vv}$ is the vector of unstructured random effects and ${\uu}_{\star}$ is the scaled intrinsic CAR model with generalized variance equal to one. Note that the variance of ${\xxi}$ is expressed as a weighted average of the covariance matrices of the unstructured and structured spatial components (unlike the CAR model proposed by \cite{leroux1999estimation} which considers a weighted combination of the precision matrices), i.e.,
\begin{equation*}
{\xxi} \sim N({\0},{\Q}_{\xi}^{\star}), \;\; \mbox{with} \;\;
{\Q}_{\xi}^{\star} = \tau_{\xi}^{-1}[(1-\lambda_{\xi}){\I}_n + \lambda_{\xi}{\R}_{\star}^{-}],
\end{equation*}
where ${\I}_n$ is the $n \times n$ identity matrix, and ${\R}_{\star}^{-}$ indicates the generalised inverse of the scaled spatial structure matrix corresponding to the undirected graph of the regions under study (see, e.g., \citealp{sorbye2014scaling}).
Recall that the spatial structure matrix is defined as ${\R}_{\xi}={\D}_{W}-{\W}$, where ${\D}_{W}=diag(w_{1+},\ldots,w_{n+})$ and $w_{i+}=\sum_j w_{ij}$ is the $i^{th}$ row sum of the binary adjacency matrix ${\W}=(w_{ij})$, whose $ij^{th}$ element is equal to one if areas $i$ and $j$ are defined as neighbours (usually if they share a common border), and it is zero otherwise.

For the temporally structured random effect ${\ggamma}=(\gamma_1,\ldots,\gamma_T)^{'}$, random walks of first (RW1) or second order (RW2) prior distributions can be assumed as follow
\begin{equation*}
{\ggamma} \sim N({\0},[\tau_{\gamma}{\R}_{\gamma}]^{-}),
\end{equation*}
where $\tau_{\gamma}$ is a precision parameter and ${\R}_{\gamma}$ is the $T \times T$ structure matrix of a RW1/RW2 (see \cite{rue2005gaussian}, pp. 95 and 110).

Finally, for the space-time interaction random effect ${\ddelta}=(\delta_{11},\ldots,\delta_{n1},\ldots,\delta_{1T},\ldots,\delta_{nT})^{'}$ the following prior distribution is assumed
\begin{equation*}
{\ddelta} \sim N({\0},[\tau_{\delta}{\R}_{\delta}]^{-}),
\end{equation*}
where $\tau_{\delta}$ is a precision parameter and ${\R}_{\delta}$ is the $nT \times nT$ matrix obtained as the Kronecker product of the corresponding spatial and temporal structure matrices, where four different types of interactions where originally proposed by \cite{KnorrHeld2000} (see \autoref{tab:ST_interactions}).

\begin{table}[!ht]
\caption{Specification for different types of space-time interactions. \label{tab:ST_interactions}}
\begin{center}
\renewcommand{\arraystretch}{1}
\begin{tabular}{lccc}
\hline
\multirow{2}{*}{Interaction} & \multirow{2}{*}{${\R}_{\delta}$} & Spatial & Temporal \\[-0.5ex]
& & correlation & correlation \\
\hline
Type I   & ${\I}_T \otimes {\I}_n$ & $-$ & $-$ \\
Type II  & ${\R}_{\gamma} \otimes {\I}_n$ & $-$ & \checkmark \\
Type III & ${\I}_T \otimes {\R}_{\xi}$ & \checkmark & $-$ \\
Type IV  & ${\R}_{\gamma} \otimes {\R}_{\xi}$ & \checkmark & \checkmark \\
\hline
\end{tabular}
\end{center}
\end{table}

In what follows, we will refer to Model \ref{eq:Global_model} as the \textit{Global model}. These models are flexible enough to describe many real situations, and their interpretation is simple and attractive. However, the models are typically not identifiable and appropriate sum-to-zero constraints must be imposed over the random effects \citep{goicoa2018spatio}. See \autoref{tab:Constraints} for a full description of the identifiability constraints that need to be imposed on each type of space-time interaction.

\begin{table}[!t]
\caption{Identifiability constraints for the different types of space-time interaction effects in CAR models \citep{goicoa2018spatio}. \label{tab:Constraints}}
\begin{center}
\renewcommand{\arraystretch}{1.5}
\normalsize
\resizebox{\textwidth}{!}{
\begin{tabular}{lcc}
\hline
Interaction & ${\R}_{\delta}$ & Constraints \\
\hline
Type I   & ${\I}_T \otimes {\I}_n$ & $\sum\limits_{i=1}^S \xi_i=0, \, \sum\limits_{t=1}^T \gamma_t=0, \, \mbox{ and } \, \sum\limits_{i=1}^S \sum\limits_{t=1}^T \delta_{it}=0$ \\
Type II  & ${\R}_{\gamma} \otimes {\I}_n$ & $\sum\limits_{i=1}^S \xi_i=0, \, \sum\limits_{t=1}^T \gamma_t=0, \, \mbox{ and } \, \sum\limits_{t=1}^T \delta_{it}=0, \, \mbox{for } \, i=1,\ldots,S$ \\
Type III & ${\I}_T \otimes {\R}_{\xi}$ & $\sum\limits_{i=1}^S \xi_i=0, \, \sum\limits_{t=1}^T \gamma_t=0, \, \mbox{ and } \, \sum\limits_{i=1}^S \delta_{it}=0, \, \mbox{for } \, t=1,\ldots,T$ \\
Type IV  & ${\R}_{\gamma} \otimes {\R}_{\xi}$ & $\sum\limits_{i=1}^S \xi_i=0, \, \sum\limits_{t=1}^T \gamma_t=0, \,  \mbox{ and } \,
\begin{array}{l} \sum\limits_{t=1}^T \delta_{it}=0, \, \mbox{for } \, i=1,\ldots,S, \\ \sum\limits_{i=1}^S \delta_{it}=0, \, \mbox{for } \, t=1,\ldots,T. \\ \end{array}$\\[5.ex]
\hline
\end{tabular}}
\end{center}
\end{table}

\subsection{Model fitting via integrated nested Laplace approximations}
Bayesian inference has traditionally been used to fit spatial and spatio-temporal disease mapping models. The fully Bayesian approach provides posterior distributions of model parameters instead of a single point estimate. However, these distributions cannot usually be derived analytically and simulation techniques based on Markov chain Monte Carlo (MCMC) methods have been traditionally used for Bayesian inference \citep{gilks1995markov}. Although these simulation-based techniques are widely used, mainly due to the development of free software to run MCMC algorithms such as WinBUGS \citep{spiegelhalter2003winbugs}, JAGS \citep{plummer2003jags}, STAN \citep{carpenter2017stan} or NIMBLE \citep{de2017programming}, these methods tend to be computationally very demanding and large Monte Carlo errors are usually present for complex spatio-temporal models \citep{schrodle2011using}. An alternative method to improve the speed of these calculations is to approximate the marginal posteriors of the model parameters using integrated nested Laplace approximations (INLA) \citep{rue2009approximate}. The INLA technique is especially attractive for latent GMRFs with sparse precision matrices and is being increasingly used in applied statistics in general \citep{rue2017bayesian} and in the field of spatial statistics in particular \citep{bakka2018spatial}. Recently, NIMBLE and R-INLA have been compared in a simulation study to fit spatio-temporal disease mapping models \citep{urdangarin2022}. The results obtained are identical in terms of relative risk estimates and nearly identical in terms of parameter estimates. However, R-INLA is considerably faster than NIMBLE.

\section{Methodology}
\label{sec:Scalable_models}

There is no doubt that the use of spatio-temporal CAR models allows to obtain accurate risk estimates in reasonable computational times when the number of areal-units is relatively small. However, two main issues arise when analyzing very large spatio-temporal datasets: (i) computational time and resources, and (ii) model assumptions. Most of the smoothing methods proposed in the literature (including CAR models) are built on the idea of spatial/temporal correlation and generally use a covariance or precision matrix with dimension equal to the number of data (spatial locations $\times$ time points), leading to prohibitive computational times if (partial) matrix inversions are necessary during the estimation process. In addition, CAR models induce the same degree of spatial dependence through the whole adjacency graph. However, the larger a spatial domain is, the less likely is that the data are stationary across the whole map.

With the objective of overcoming these problematic aspects, we propose a scalable and non-stationary Bayesian modelling approach by extending the spatial models described in \cite{orozco2021scalable} based on the idea of ``divide-and-conquer'' so that local spatio-temporal models can be simultaneously fitted. Our modelling approach consists of three main steps. First, the region of interest is divided into $D$ subdomains. Then, local spatio-temporal models are fitted using a fully Bayesian approach based on INLA. Finally, the results are merged to obtain posterior marginal estimates of the relative risks for each areal-time unit. Instead of considering global random effects whose correlation structures are based on the whole spatial/temporal neighbourhood graphs of the areal-units, as is the case of the Global model described in Equation \eqref{eq:Global_model}, we propose to divide the data into subdomains based on spatial partitions so that models with different local correlation structures, that is, models inducing different amount of smoothing are defined. Then, extending the methodology described in \cite{orozco2021scalable}, \textit{Disjoint} and \textit{k-order neighbourhood models} are defined for estimating spatio-temporal disease risks.

For the \textit{Disjoint model}, a partition of the spatial domain $\mathscr{D}$ into $D$ subdomains is defined, so that $\mathscr{D} = \bigcup_{d=1}^D \mathscr{D}_d$ where $\mathscr{D}_j \cap \mathscr{D}_k = \emptyset$ for all $j \neq k$. If we denote as $A_{it}$ to the small area $i$ in time period $t$, let ${\bf O}_{d}=\{O_{it} \, | \, A_{it} \in \mathscr{D}_d\}$ and ${\bf E}_{d}=\{E_{it} \, | \, A_{it} \in \mathscr{D}_d\}$ represent the observed and expected number of disease cases in each subdomain, respectively. Then, $D$ independent local spatio-temporal models similar to those described in Section~\ref{sec:ST_models} are simultaneously fitted. Since each areal-time unit $A_{it}$ belongs to a single subdomain, the final log-risk surface $\log{\rr}=(\log{\rr}_1,\ldots,\log{\rr}_D)^{'}$ is just the union of the posterior marginal estimates of each spatio-temporal sub-model.

\begin{figure}[!ht]
\begin{center}
\vspace{0.5cm}
\includegraphics[width=\textwidth]{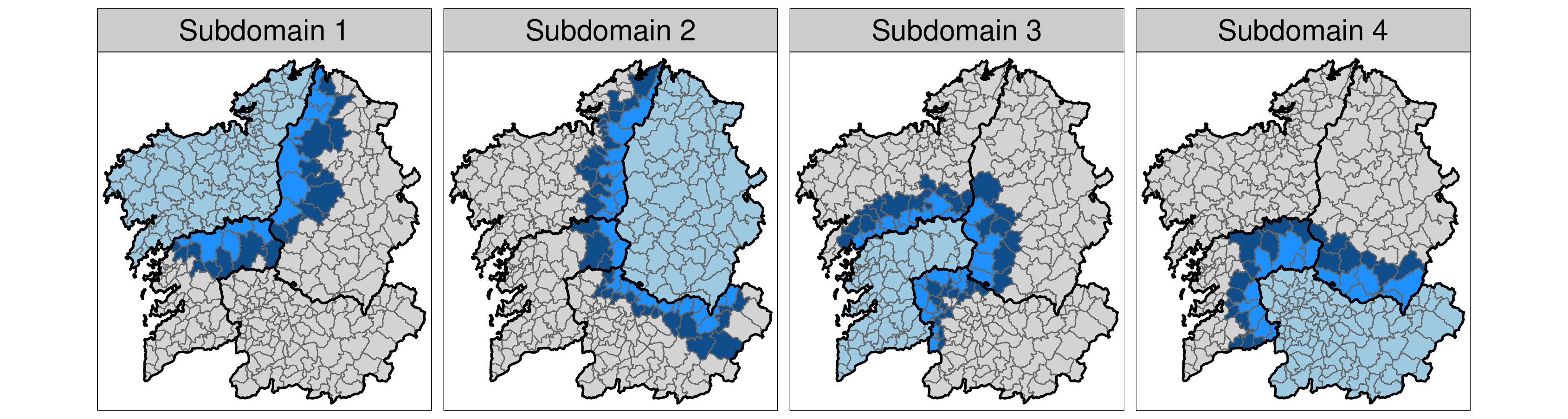}\hfill
\vspace{-0.5cm}
\end{center}
\caption{Toy example of a spatial partition into $D=4$ subdomains. Light-blue areas represents those corresponding to the Disjoint models, while spatial adjacent areas are added when considering the 1st/2nd-order neighbourhood models (blue and dark-blue areas, respectively). \label{fig:Example_Subdomains}}
\end{figure}

However, assuming independence between areal-time units belonging to different subdomains could be very restrictive and may lead to border effects. To avoid this undesirable issue, we define the \textit{k-order neighbourhood model} by adding neighbouring areal units (based on spatial adjacency) to each partition. A toy example of a spatial partition into four subdomains is represented in \autoref{fig:Example_Subdomains}. Since multiple log-risk estimates can be obtained for some $A_{it}$ units from the different local submodels, their posterior estimates must be properly combined to obtain a single posterior distribution for each $r_{it}$. Originally, \cite{orozco2021scalable} propose to compute mixture distributions of the estimated posterior probability density functions with weights proportional to the conditional predictive ordinates (CPOs; \citealp{pettit1990conditional}). Here, we also investigate the strategy of using the posterior marginal risk estimates of $A_{it}$ corresponding to the original domain the $i$-th area belonged to. A full comparison in terms of risk estimation accuracy and high/low risk area detection using these two merging strategies is described in Section~\ref{sec:SimulationStudy}.

\section{Software: Model implementation in the \texttt{R} package \texttt{bigDM}}
\label{sec:bigDM}

We have implemented several scalable spatio-temporal disease mapping models in the \texttt{R} package \texttt{bigDM} \citep{bigDM}
(see \url{https://github.com/spatialstatisticsupna/bigDM}). 
A generic scheme of the main algorithms 
is described in Algorithms~\ref{alg:STCAR_INLA}, \ref{alg:mixtures} and \ref{alg:DIC}.
Since in the disease mapping context the high-dimensionality of the data is usually related to a large number of small areas, we consider only purely spatial partitions. These partitions could be based on administrative divisions of the area of interest (such as provinces, states or local health areas), or random partitions based on a regular grid over the associated cartography. However, random partitions should be carefully done, since small domains with large number of areas with no observed cases could lead to wrong model estimates.

When fitting both the Disjoint and k-order neighbourhood models, parallel or distributed computation strategies can be performed to speed up computations by using the \texttt{future} package \citep{bengtsson2020unifying}. If the \sloppy \texttt{plan="sequential"} argument is specified, the models are fitted one at a time in the current R session (local machine). In contrast, if the \texttt{plan="cluster"} argument is defined, multiple models can be fitted in parallel on external R sessions (local machine) or distributed in remote computing nodes. When using this option, the identifications of the local/remote workers where the models are going to be processed must be configured through the \texttt{workers} argument. As it is well known, the communication between the ``master node'' and the rest of workers affects the computational time, so the decision on how to configure the processing architecture must be made carefully (depending on the characteristics of the computations to be performed).

As described in the previous section, two different merging strategies could be considered to properly combined the posterior marginal estimates of the relative risks when fitting the k-order neighbourhood models. If the \texttt{merge.strategy="mixture"} argument is specified, mixture distributions of the posterior probability density functions are computed for each $\log{r_{it}}$ (see Algorithms~\ref{alg:mixtures}). On the other hand, if the \texttt{merge.strategy="original"} argument is specified, the posterior marginal estimate of the areal-unit corresponding to the original subdomain is selected.

In addition, approximations to model selection criteria such as the deviance information criterion (DIC) \citep{spiegelhalter2002bayesian} and the Watanabe-Akaike information criterion (WAIC) \citep{watanabe2010asymptotic}, two widely used criteria to compare models in a fully Bayesian setting, are also derived by default when fitting the scalable models using the \texttt{bigDM} package.
Specific vignettes accompanying the package have been included to facilitate the use of the methodology for non-expert users (see  \url{https://emi-sstcdapp.unavarra.es/bigDM/bigDM-3-fitting-spatio-temporal-models.html}).


\smallskip

\begin{algorithm*}[!ht]
\caption{Fit a scalable spatio-temporal model for high-dimensional areal count data.\label{alg:STCAR_INLA}}
{\fontsize{11}{11}\selectfont
\vspace{0.1cm}
{\bf Inputs:}
    \vspace{-0.3cm}
    \begin{itemize}
    \itemsep-0.4em
	\item Cartography file with count data corresponding to areal units $A_{it}$, for $i=1,\ldots,n$, and $t=1,\ldots,T$.
    \item Observed cases $O_{it}$ and expected cases $E_{it}$.
    \item Prior distributions for the spatial ($\xxi$), temporal ($\ggamma$) and spatio-temporal ($\ddelta$) random effects.
    \item \texttt{W}: binary adjacency matrix of the spatial areal units.
    \item \texttt{k}: numeric value with the neighbourhood order.
    \item \texttt{plan}: computation strategy used for model fitting (one of either \texttt{"sequential"} or \texttt{"cluster"}).
    \item \texttt{workers}: IDs of the local or remote workers (only required if \texttt{plan="cluster"}).
    \item \texttt{merge.strategy}: merging strategy to compute posterior marginal estimates of relative risks. One of either \texttt{"mixture"} or \texttt{"original"} (default).
    \end{itemize}
{\bf Step 1: Pre-processing the data}
    \begin{algorithmic}[1]
    \If{\texttt{W=NULL}}
        \State compute \texttt{W} from the cartography file.
    \EndIf
    \State Merge disjoint connected subgraphs.
    \State Define \texttt{formula} object for INLA model according to the prior distributions for $\xxi$, $\ggamma$ and $\ddelta$.
    \end{algorithmic}
\vspace{0.1cm}
{\bf Step 2: Fitting submodels with INLA}
    \begin{algorithmic}[1]
    \State Divide the spatial domain into $D$ subdomains.
    \For{$d \in \{1,\ldots,D\}$}
        \If{\texttt{k>0}}
            \State add $k$-order neighbouring areas.
        \EndIf
		\State Compute the spatial adjacency matrix $W_{d}$.
		\State Extract ${\bf O}_d=\{O_{it} | A_{it} \in \mathscr{D}_d \}$ and ${\bf E}_d=\{E_{it} | A_{it} \in \mathscr{D}_d \}$.
        \State Define appropriate identifiability constraints.
        \If{\texttt{plan="sequential"}}
            \State fit INLA models sequentially in the current R session (local machine).
        \Else
            \State fit INLA models in parallel on external R sessions (local machine) or distributed in remote compute nodes.
        \EndIf
	\EndFor
    \end{algorithmic}
\vspace{0.1cm}
{\bf Step 3: Merging results}
    \begin{algorithmic}[1]
    \If{\texttt{plan="cluster"}}
        \State retrieve submodels to the central node.
    \EndIf
    \If{\texttt{k>0} and \texttt{merge.strategy="mixture"}}
            \State compute mixture distributions for the posterior probability density functions of each $\log{r_{it}}$. \Comment{Algorithm 2}
    \EndIf
    \If{\texttt{k>0} and \texttt{merge.strategy="original"}}
            \State select the posterior marginal estimate of the areal-unit corresponding to the original subdomain
    \EndIf
    \State Compute approximate DIC and WAIC values. \Comment{Algorithm 3}
    \end{algorithmic}
\vspace{0.1cm}
{\bf Output:}
    \vspace{-0.3cm}
    \begin{itemize}
	\item \texttt{inla} object with the fitted model.
    \end{itemize}
}
\end{algorithm*}

\clearpage

\begin{algorithm*}[!ht]
\caption{Compute mixture distributions for each $\log{r_{it}}$.\label{alg:mixtures}}
\vspace{0.1cm}
{\fontsize{11}{11}\selectfont
{\bf Inputs:} \texttt{inla} submodel $d \in \{1,\ldots,D\}$ containing
    \vspace{-0.2cm}
    \begin{itemize}
    \itemsep0em
	\item $f_d(x)$: posterior probability density function estimates of the log-risk for areal-unit $A_{it}$.
    \item $CPO_{it}^d=Pr(O_{it}=o_{it}|\mathbf{o}_{-it})$: conditional predictive ordinate for areal-unit $A_{it}$.
    \item \texttt{p}: number of equally spaced points at which the density is evaluated (default to 75).
    \end{itemize}
{\bf Parallel computation of mixture distributions}
    \begin{algorithmic}[1]
    \For{$i \in \{1,\ldots,n\}$ and $t \in \{1,\ldots,T\}$}
        \State Compute $m(i)$: number of submodels in which the areal-unit $A_{it}$ has been estimated (note that $m(i)<D$)
        \State Compute normalized weights $$w_j=\dfrac{CPO_{it}^j}{\sum\limits_j CPO_{it}^j}, \quad \mbox{for} \quad j=1,\ldots,m(i).$$
        \vspace{-0.5cm}
        \State Compute $f(x)=\sum\limits_{j=1}^{m(i)} w_j f_j(x)$ evaluated at $p$ points.
	\EndFor
    \end{algorithmic}
{\bf Output:}
    \vspace{-0.2cm}
    \begin{itemize}
	\item Posterior marginal density estimates of log-risks.
    \end{itemize}
}
\end{algorithm*}

\begin{algorithm*}[!ht]
\caption{Computations of approximate DIC and WAIC values.\label{alg:DIC}}
{\fontsize{11}{11}\selectfont
\vspace{0.1cm}
{\bf Inputs:}
    \vspace{-0.2cm}
    \begin{itemize}
    \itemsep0em
	\item Posterior marginal density estimates of the risks for each $A_{it}$.
    \item $S$: number of samples to draw (default to 1000).
    \end{itemize}
\vspace{0.1cm}
{\bf Parallel computation:}
    \begin{algorithmic}[1]
    \For{$i \in \{1,\ldots,n\}$ and $t \in \{1,\ldots,T\}$}
        \State Draw $S$ samples from the posterior marginal distribution of $r_{it}$.
        \State Compute ${\ttheta}^s$: posterior simulations of $\mu_{it}=E_{it}r_{it}$.
        \State Compute the deviance information criterion $\mbox{DIC} = 2\overline{D({\ttheta})}-D(\bar{\ttheta})$, by approximating the mean deviance $\overline{D({\ttheta})}$ and the deviance of the mean $D(\bar{\ttheta})$ as
        $$\overline{D({\ttheta})} \approx \frac{1}{S} \sum\limits_{s=1}^S -2 \log(p({\bf O}|{\ttheta}^s)), \qquad
        D(\bar{\ttheta}) \approx -2 \log(p({\bf O}|\bar{\ttheta})), \quad \mbox{with } \bar{\ttheta}=\frac{1}{S}\sum\limits_{s=1}^S {\ttheta}^s,$$ where $p({\bf O}|\ttheta)$ is the likelihood function of a Poisson distribution with mean $\ttheta$.
        \State Approximate Watanabe-Akaike information criterion as
        $$\mbox{WAIC}=-2 \sum\limits_{i=1}^{n}\sum\limits_{t=1}^{T}\log \left(\frac{1}{S}\sum\limits_{s=1}^{S} p(O_{it}|{\ttheta}^s)\right) + 2 \sum\limits_{i=1}^{n}\sum\limits_{t=1}^{T} \mbox{Var}\left[ \log(p({O_{it}}|{\ttheta}^s)) \right]$$
	\EndFor
    \end{algorithmic}
\vspace{-0.5cm}
{\bf Output:}
    \vspace{-0.2cm}
    \begin{itemize}
	\item Approximate DIC and WAIC values.
    \end{itemize}
}
\end{algorithm*}

\section{Simulation study}
\label{sec:SimulationStudy}

In this section, we present two simulation studies. The first one compares the performance of our scalable model proposals with the commonly used disease mapping models described in Section~\ref{sec:ST_models} (denoted as Global models) in terms of risk estimation accuracy and high/low risk area detection. The second one evaluates the computing speed offered by our modelling approach when using both parallel and/or distributed computation strategies as the number of small areas increases.

\subsection{Risk estimation in high-dimensional areal data}
\label{sec:SimulationStudy1}

\subsubsection{Data generation}
\label{sec:SimulationStudy_Data}
\begin{figure}[!ht]
\begin{center}
\includegraphics[width=1.1\textwidth]{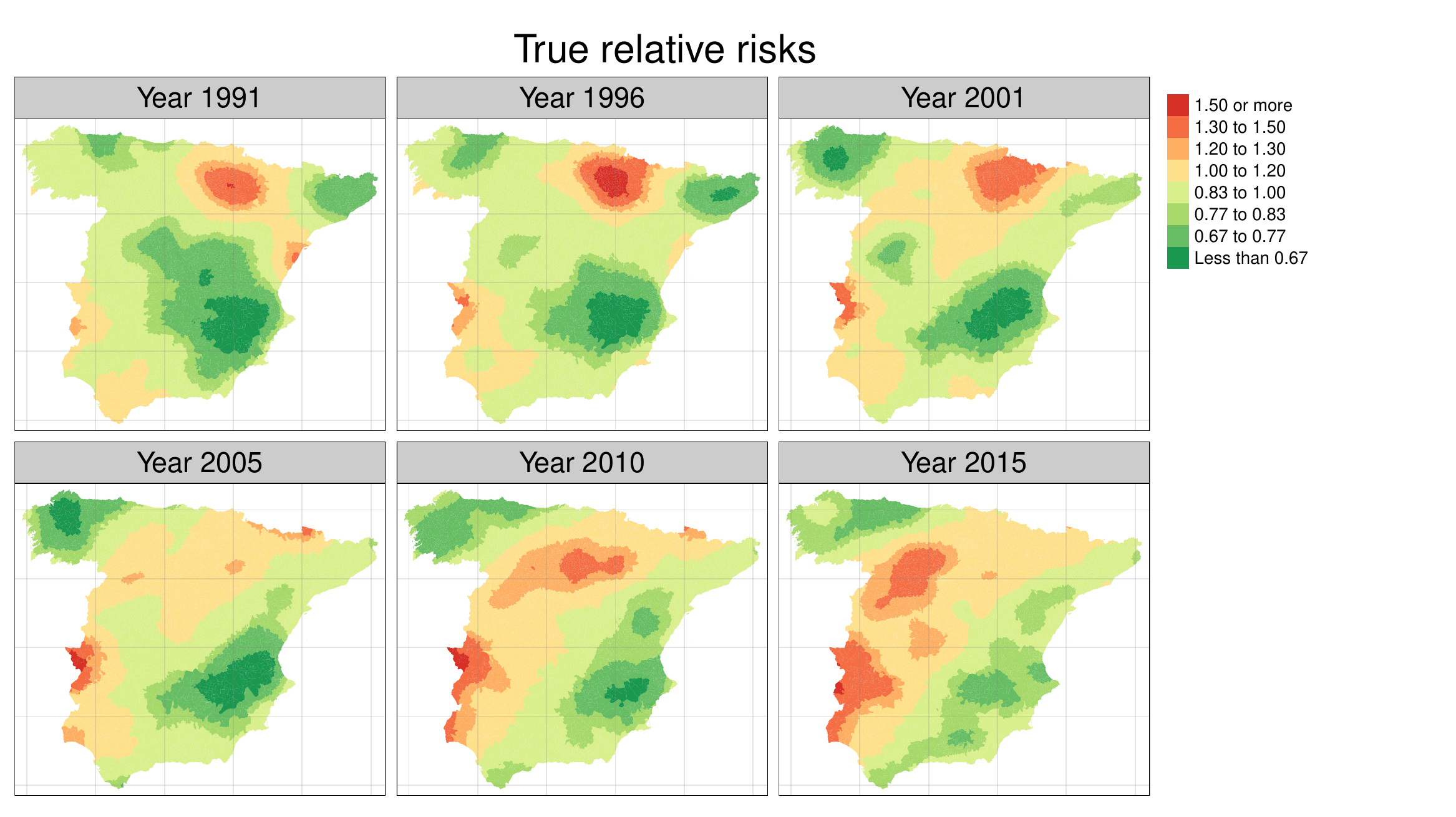}\hfill
\includegraphics[width=1.1\textwidth]{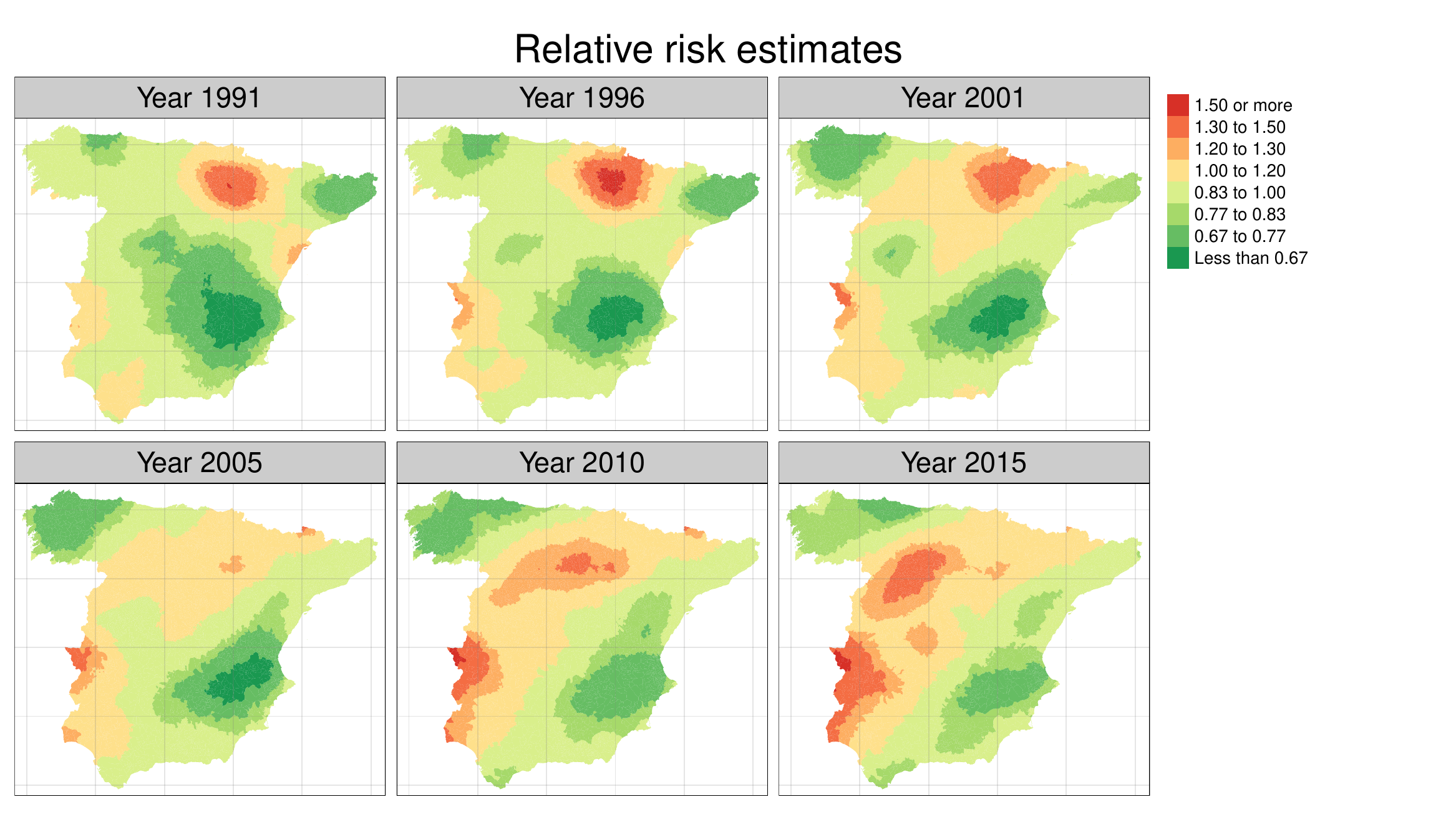}\hfill
\vspace{-1cm}
\end{center}
\caption{Simulation study: true risk surfaces (top) and average values of posterior median estimates of relative risks for 2nd-order neighbourhood and Type IV interaction model using the \texttt{"original"} merging strategy (bottom) for some selected years. \label{fig:True_risks}}
\end{figure}

The $n=7907$ municipalities of continental Spain and $T=25$ time periods are used as the simulation template. For the scalable model proposals, we divide the data into $D=47$ subdomains using the provinces of Spain to define a spatial partition, as this is the setting for the real data analysis presented in the next section. Under this template, a smooth risk surface is generated by sampling from a three-dimensional P-spline with 20 equally spaced knots for longitude and latitude, and 6 equally spaced knots for time. The true risk surfaces for the simulation study are shown on top of \autoref{fig:True_risks}. The simulated counts for each municipality and time point are generated from a Poisson distribution with mean $E_{it}r_{it}$, where the number of expected cases $E_{it}$ are fixed at value 10. A total of 50 simulated datasets have been generated.

\subsubsection{Results}
\label{sec:SimulationStudy_Results}
Four different spatio-temporal models have been fitted to each simulated dataset: the Global model (Section~\ref{sec:ST_models}) and the Disjoint, 1st-order neighbourhood and 2nd-order neighbourhood scalable model proposals (Section~\ref{sec:Scalable_models}). For all these models, a BYM2 prior for the spatial random effect, a RW1 prior for the temporal random effect and the four types of interactions originally proposed by \cite{KnorrHeld2000} for the spatio-temporal random effect have been considered.

Regarding model hyperparameters, improper uniform prior distributions are given to all the standard deviations (inverse square root of the precision parameters), and uniform prior distributions on the interval $[0,1]$ are given to the spatial smoothing parameters of the BYM2 prior. Finally, a vague zero mean normal distribution with a precision close to zero (0.001) is given to the model intercept. 
All the calculations are made on a single machine with a Intel Xeon E5-2620 v4 processor and 256GB RAM (CentOS Linux release 7.3.1611 operative system), using the Gaussian approximation strategy in R-INLA (stable version INLA\_22.05.07) of R-4.1.3 and simultaneously running 12 models in parallel using the \texttt{bigDM} package.

\clearpage

We evaluate the model performance in terms of how well relative risk are estimated by computing the mean absolute relative bias (MARB) and mean relative root mean square error (MRRMSE) for each municipality defined as
\begin{eqnarray*}
\begin{array}{rcl}
\mbox{MARB}_i & = & \frac{1}{T}\sum\limits_{t=1}^T \frac{1}{50} \left|\sum\limits_{l=1}^{50} \frac{\hat{r}_{it}^l-r_{it}}{r_{it}} \right|,
\\[3.ex]
\mbox{MRRMSE}_i & = & \frac{1}{T}\sum\limits_{t=1}^T \sqrt{\frac{1}{50} \sum\limits_{l=1}^{50} \left(\frac{\hat{r}_{it}^l-r_{it}}{r_{it}} \right)^2},
\end{array}
\end{eqnarray*}
where $r_{it}$ is the true generated risk, and $\hat{r}_{it}^l$ is the posterior median estimate of the relative risk for areal unit $i$ and time period $t$ in the $l$-th simulation.
We also compute the Interval Score (IS) for the 95\% credible interval of the risks, a proper scoring rule for quantile predictions (see e.g., \citealp{gneiting2007strictly}) that combines both the length and the empirical coverage of the credible interval which is defined as
\begin{eqnarray*}
IS_{0.05}(r) = (u-l)+\dfrac{2}{0.05}(l-r)I[r<l]+ +\dfrac{2}{0.05}(r-u)I[r>u],
\end{eqnarray*}
where $[l,u]$ is the 95\% credible interval for the risk and $I[\cdot]$ denotes an indicator function that penalizes the length of the credible interval if the real risk ($r$) is not contained within that interval. For all these criteria, lower values imply better model properties.

 The results of the simulation study are summarized in \autoref{tab:SimulationStudy}, where average values of Bayesian model selection criteria, risk estimation accuracy measures and computational time are displayed. For the 1st/2nd-order neighbourhood models, we compare both \texttt{"mixture"} and \texttt{"original"} merging strategies.
We notice that it was computationally unfeasible to fit Type II and Type IV interaction Global models, because of the huge dimension of the spatio-temporal structure matrix ($\approx 4\times10^{10}$ elements) and the high number of identifiability constraints over the spatio-temporal interaction ($\approx 8000$ constraints). In contrast, we were able to fit our scalable model proposals reducing the RAM/CPU memory usage and computational time substantially. The computational time for the Disjoint and k-order neighbourhood models are divided into \textit{running time} and \textit{merging time}. For the Global models only the running time is computed (we remind that these models are not scalable). The running time refers to the elapsed time for all the submodels (which can be fitted in both parallel and/or distributed processing architectures) and the merging time corresponds to the computation of how the posterior distribution of the log-risks are combined (when necessary) and computation of approximate DIC/WAIC values in the ``master node''. As expected, the complexity and computational time of the models increase as higher values of neighbourhood order are considered. On the contrary, the merging time only increases as higher neighbourhood order models are considered, as the number of areal-units for which posterior estimates must be combined increases. As it is shown in \autoref{tab:SimulationStudy}, the \texttt{"original"} merging strategy is less computational demanding than using mixture distributions.

\begin{table}[!ht]
\caption{Simulation study: average values of mean deviance $\overline{{D(\theta)}}$, effective number of parameters ($p_D$), deviance information criterion (DIC), Watanabe-Akaike information criterion (WAIC), mean absolute relative bias (MARB), mean relative root mean square error (MRRMSE), Interval Score (IS) and computational time in minutes (T.run: running time, T.merge: merging time). For the 1st/2nd-order neighbourhood models, both \texttt{"mixture"} and \texttt{"original"} merging strategies are compared. \label{tab:SimulationStudy}}
\vspace{-0.5cm}
\renewcommand{\arraystretch}{1.2}
\begin{center}
\resizebox{\textwidth}{!}{
\begin{tabular}{llrrrrrrrrrrrr}
\hline\noalign{\smallskip}
& & \multicolumn{4}{c}{Model selection criteria} & & \multicolumn{3}{c}{Risk estimation accuracy} & & \multicolumn{2}{c}{Time} \\
\cline{3-6} \cline{8-10} \cline{12-13} \\
Model & Interaction & $\overline{{D(\theta)}}$ & $p_D$ & DIC & WAIC & & MARB & MRRMSE & IS & & T.run & T.merge \\
\noalign{\smallskip}\hline\noalign{\smallskip}
Global        & Type I   & 201406 & 15987 & 217393 & 217832 & & 0.0684 & 0.0782 & 0.3959 & & 14  &     \\
              & Type II  &    $-$ &   $-$ &    $-$ &    $-$ & &    $-$ &    $-$ &    $-$ & & $-$ & $-$ \\
              & Type III & 194247 & 10683 & 204930 & 204666 & & 0.0165 & 0.0387 & 0.2816 & & 301 &     \\
              & Type IV  &    $-$ &   $-$ &    $-$ &    $-$ & &    $-$ &    $-$ &    $-$ & & $-$ & $-$ \\
\noalign{\smallskip}\hline\noalign{\smallskip}
Disjoint      & Type I   & 198972 & 7563 & 206536 & 206516 & & 0.0322 & 0.0434 & 0.2582 & & 3  & 6 \\
              & Type II  & 200225 & 5710 & 205934 & 205965 & & 0.0281 & 0.0419 & 0.2231 & & 34 & 6 \\
              & Type III & 197112 & 7817 & 204929 & 204829 & & 0.0203 & 0.0377 & 0.2404 & & 7  & 6 \\
              & Type IV  & 199103 & 5059 & 204162 & 204151 & & 0.0153 & 0.0331 & 0.1950 & & 64 & 6 \\
\noalign{\smallskip}\hline\noalign{\smallskip}
& & \multicolumn{9}{c}{\bf merge.strategy=``mixture"} \\
\noalign{\smallskip}\hline\noalign{\smallskip}
1st order     & Type I   & 197900 & 8104 & 206005 & 205948 & & 0.0303 & 0.0416 & 0.2550 & & 3  & 18 \\
neighbourhood & Type II  & 199211 & 6320 & 205531 & 205534 & & 0.0261 & 0.0404 & 0.2239 & & 53 & 18 \\
              & Type III & 196069 & 8366 & 204434 & 204297 & & 0.0173 & 0.0352 & 0.2490 & & 10 & 18 \\
              & Type IV  & 198603 & 5243 & 203846 & 203823 & & 0.0133 & 0.0311 & 0.1974 & & 70 & 18 \\
\noalign{\smallskip}\hline\noalign{\smallskip}
2nd order     & Type I   & 197325 & 8821 & 206146 & 206073 & & 0.0312 & 0.0423 & 0.2614 & & 4   & 32 \\
neighbourhood & Type II  & 198631 & 7070 & 205701 & 205689 & & 0.0266 & 0.0413 & 0.2341 & & 124 & 32 \\
              & Type III & 195532 & 8986 & 204518 & 204353 & & 0.0173 & 0.0356 & 0.2593 & & 16  & 32 \\
              & Type IV  & 198477 & 5397 & 203874 & 203848 & & 0.0134 & 0.0312 & 0.2008 & & 136 & 32 \\
\noalign{\smallskip}\hline\noalign{\smallskip}
& & \multicolumn{9}{c}{\bf merge.strategy``original"} \\
\noalign{\smallskip}\hline\noalign{\smallskip}
1st order     & Type I   & 198937 & 7435 & 206373 & 206356 & & 0.0322 & 0.0427 & 0.2529 & & 3  & 8 \\
neighbourhood & Type II  & 199954 & 5836 & 205790 & 205813 & & 0.0276 & 0.0414 & 0.2215 & & 53 & 8 \\
              & Type III & 196811 & 7712 & 204523 & 204424 & & 0.0182 & 0.0357 & 0.2395 & & 10 & 8 \\
              & Type IV  & 198976 & 4885 & 203861 & 203846 & & 0.0137 & 0.0312 & 0.1911 & & 70 & 8 \\
\noalign{\smallskip}\hline\noalign{\smallskip}
2nd order     & Type I   & 198881 & 7592 & 206472 & 206461 & & 0.0332 & 0.0432 & 0.2524 & & 4   & 9 \\
neighbourhood & Type II  & 199640 & 6247 & 205887 & 205905 & & 0.0278 & 0.0421 & 0.2259 & & 124 & 9 \\
              & Type III & 196473 & 7891 & 204364 & 204253 & & 0.0173 & 0.0348 & 0.2419 & & 16  & 9 \\
              & Type IV  & 198897 & 4870 & {\bf 203767} & {\bf 203751} & & {\bf 0.0133} & {\bf 0.0305} & {\bf 0.1903} & & 136 & 9 \\
\noalign{\smallskip}\hline
\end{tabular}}
\end{center}
\end{table}

Models with a completely structured space-time interaction (Type IV interaction) perform better in this scenario in terms of model selection criteria and risk estimation accuracy measures, followed by Type III interaction models. The main reason could be that in the true risk surface considered in this scenario, the spatial pattern variability is greater than the temporal one. Specifically, the percentages of variability of the overall risk explained by each pattern is about 70\% spatial, 5\% temporal and 25\% spatio-temporal.
If the \texttt{"mixture"} strategy is used to combine the posterior marginal estimates of the relative risks in the border areas (that is, areal-units that are estimated in more than one submodel), slightly better results are obtained with 1st-order neighbourhood models in comparison with those considering 2nd-order neighbours. However, if the \texttt{"original"} strategy is used to combine the estimated risks from the different submodels, 2nd-order neighbourhood models performs better for Type III and Type IV interactions. As expected, the differences between models are more clearly stated if we compute these measures only for those areal-units located in the borders of the partition of the spatial domain (see \autoref{tab:Appendix1}). As it is shown at the bottom of \autoref{fig:True_risks}, the average values of posterior median estimates of relative risks obtained with 2nd-order neighbourhood and Type IV interaction model using the \texttt{"original"} merging strategy are very similar to the true risk surface.

We are also interested in evaluating the models in terms of their ability to detect true high and low risk areas by calculating true positive/negative rates and false positive/negative rates. For each areal-time unit $A_{it}$, a high (low) risk area is an area where the true risk $r_{it}$ is greater (less) than one. After fitting the model, we classify an area as having high risk if the posterior probabilities that $r_{it}$ exceeds 1 is higher than a threshold value $p_0$, namely the exceedence probabilities $P(r_{it}>1 | {\bf O})>p_0$. Conversely, we classify a low risk area if the posterior probabilities that $r_{it}$ is below 1 is higher than $p_0$, i.e., $P(r_{it}<1 | {\bf O})>p_0$. Notice that these probabilities are computed from the posterior marginal distributions of the estimated relative risks.
True positive rates (TPR or sensitivity) are computed as the proportion of high true risks ($r_{it}>1$) that were correctly classified as a high risk area, while true negative rates (TNR or specificity) are computed as the proportion of low true risks ($r_{it}<1$) that were correctly classified as a low risk area. At the same time, we are also interested in comparing the misclassification errors of the models in terms of false positive rates (FPR), i.e., the proportion of areas that are incorrectly classified as a high risk area, and false negative rates (FNR), i.e., the proportion of areas that are incorrectly classified as a low risk area.

Average values of TPR, FPR, TNR and FNR for the reference threshold values of $p_0=0.8, 0.9 \mbox{ and } 0.95$ are shown in \autoref{tab:SimulationStudy_TPR}. For the 1st/2nd-order neighbourhood models, both \texttt{"mixture"} and \texttt{"original"} merging strategies are compared. We notice that that our proposed scalable models outperform the Global models in terms of high and low risk area detection. In particular, the first order neighborhood model \texttt{"original"} strategy (Type IV interaction) performs the best in terms of detecting TPR. The rest of scalable models including the disjoint model (Type IV interaction) behave very similarly.
In addition, if we compute these measures only for the areal-units located in the borders of the partition of the spatial domain (see \autoref{tab:Appendix2A} and \autoref{tab:Appendix2B}) models with the \texttt{"original"} merging strategy show again better results. In general, similar values of TPR and TNR are obtained for 1st and 2nd order neighbourhood models using both merging strategies. In terms of false positive and false negative rates, although slightly better results are obtained when the \texttt{"mixture"} strategy is used, very low values are obtained in all cases.

\begin{landscape}

\begin{table}[!ht]
\caption{Simulation study: average values of true/false positive rates and true/false negative rates for the reference threshold values of $p_0=0.8, 0.9 \mbox{ and } 0.95$, based on posterior exceedence probabilities $P(r_{it}>1|{\bf O})$ and $P(r_{it}<1|{\bf O})$, respectively. For the 1st/2nd-order neighbourhood models, both \texttt{"mixture"} and \texttt{"original"} merging strategies are compared. \label{tab:SimulationStudy_TPR}}
\renewcommand{\arraystretch}{1.2}
\begin{center}
\resizebox{1.3\textwidth}{!}{
\begin{tabular}{llrrrrrrrrrrrrrrrr}
\hline\noalign{\smallskip}
& & & \multicolumn{3}{c}{True Positive Rate} & & \multicolumn{3}{c}{True Negative Rate} & & \multicolumn{3}{c}{False Positive Rate} & & \multicolumn{3}{c}{False Negative Rate}\\
\cline{4-6} \cline{8-10} \cline{12-14} \cline{16-18}\\
Model & \parbox[p][0.04\textwidth][c]{2cm}{Space-time interaction} & & $p_0=0.8$ & $p_0=0.9$ & $p_0=0.95$ & & $p_0=0.8$ & $p_0=0.9$ & $p_0=0.95$  & & $p_0=0.8$ & $p_0=0.9$ & $p_0=0.95$ & & $p_0=0.8$ & $p_0=0.9$ & $p_0=0.95$\\
\noalign{\smallskip}\hline\noalign{\smallskip}
Global     & Type I & & 0.5932 & 0.4349 & 0.3212 & & 0.6902 & 0.5453 & 0.4249 & & 0.0142 & 0.0024 & 0.0004 & & 0.0282 & 0.0130 & 0.0069\\
& Type II  & & - & - & - & & - & - & - & & - & - & - & & - & - & -\\
& Type III & & 0.7414 & 0.6211 & 0.5226 & & 0.8122 & 0.6937 & 0.5892 & & 0.0025 & 0.0003 & 0.0000 & & 0.0044 & 0.0006 & 0.0001\\
& Type IV  & & - & - & - & & - & - & - & & - & - & - & & - & - & -\\
\noalign{\smallskip}\hline\noalign{\smallskip}
Disjoint   & Type I   & & 0.7742 & 0.6706 & 0.5836 & & 0.8368 & 0.7520 & 0.6660 & & 0.0119 & 0.0043 & 0.0018 & & 0.0181 & 0.0074 & 0.0032\\
& Type II  & & 0.8021 & 0.7160 & 0.6411 & & 0.8557 & 0.7864 & 0.7209 & & 0.0111 & 0.0038 & 0.0014 & & 0.0182 & 0.0078 & 0.0036\\
& Type III & & 0.7764 & 0.6721 & 0.5868 & & 0.8403 & 0.7509 & 0.6705 & & 0.0045 & 0.0009 & 0.0002 & & 0.0080 & 0.0019 & 0.0005\\
& Type IV  & & {\bf 0.8256} & {\bf 0.7449} & {\bf 0.6754} & & {\bf 0.8747} & {\bf 0.8065} & {\bf 0.7408} & & 0.0048 & 0.0011 & 0.0003 & & 0.0079 & 0.0020 & 0.0006\\
\noalign{\smallskip}\hline\noalign{\smallskip}
& & \multicolumn{14}{c}{\bf merge.strategy="mixture"} \\
\noalign{\smallskip}\hline\noalign{\smallskip}
1st order     & Type I   & & 0.7685 & 0.6601 & 0.5688 & & 0.8345 & 0.7421 & 0.6531 & & 0.0088 & 0.0026 & 0.0009 & & 0.0148 & 0.0054 & 0.0021 \\
neighbourhood & Type II  & & 0.7941 & 0.7028 & 0.6235 & & 0.8531 & 0.7751 & 0.7039 & & 0.0082 & 0.0024 & 0.0007 & & 0.0147 & 0.0055 & 0.0023 \\
& Type III & & 0.7684 & 0.6602 & 0.5710 & & 0.8397 & 0.7413 & 0.6529 & & 0.0027 & 0.0004 & 0.0001 & & 0.0051 & 0.0009 & 0.0002 \\
& Type IV  & & {\bf 0.8208} & {\bf 0.7382} & {\bf 0.6681} & & {\bf 0.8762} & {\bf 0.8040} & {\bf 0.7349} & & 0.0033 & 0.0006 & 0.0001 & & 0.0056 & 0.0012 & 0.0003 \\
\noalign{\smallskip}\hline\noalign{\smallskip}
2nd order     & Type I   & & 0.7605 & 0.6451 & 0.5480 & & 0.8296 & 0.7272 & 0.6327 & & 0.0074 & 0.0017 & 0.0004 & & 0.0138 & 0.0046 & 0.0019 \\
neighbourhood & Type II  & & 0.7833 & 0.6860 & 0.6004 & & 0.8480 & 0.7611 & 0.6816 & & 0.0069 & 0.0018 & 0.0005 & & 0.0142 & 0.0050 & 0.0020 \\
& Type III & & 0.7614 & 0.6503 & 0.5573 & & 0.8363 & 0.7305 & 0.6347 & & 0.0024 & 0.0003 & 0.0001 & & 0.0050 & 0.0008 & 0.0002 \\
& Type IV  & & {\bf 0.8172} & {\bf 0.7336} & {\bf 0.6633} & & {\bf 0.8756} & {\bf 0.8012} & {\bf 0.7294} & & 0.0031 & 0.0006 & 0.0001 & & 0.0057 & 0.0012 & 0.0002 \\
\noalign{\smallskip}\hline\noalign{\smallskip}
& & \multicolumn{14}{c}{\bf merge.strategy="original"} \\
\noalign{\smallskip}\hline\noalign{\smallskip}
1st order     & Type I   & & 0.7736 & 0.6698 & 0.5832 & & 0.8406 & 0.7557 & 0.6699 & & 0.0111 & 0.0037 & 0.0014 & & 0.0179 & 0.0071 & 0.0029  \\
neighbourhood & Type II  & & 0.7992 & 0.7113 & 0.6354 & & 0.8573 & 0.7859 & 0.7186 & & 0.0099 & 0.0032 & 0.0010 & & 0.0174 & 0.0072 & 0.0032 \\
& Type III & & 0.7749 & 0.6707 & 0.5863 & & 0.8447 & 0.7532 & 0.6699 & & 0.0034 & 0.0006 & 0.0001 & & 0.0063 & 0.0013 & 0.0003 \\
& Type IV  & & {\bf 0.8260} & {\bf 0.7461} & {\bf 0.6782} & & {\bf 0.8798} & {\bf 0.8117} & {\bf 0.7462} & & 0.0037 & 0.0008 & 0.0002 & & 0.0064 & 0.0014 & 0.0003 \\
\noalign{\smallskip}\hline\noalign{\smallskip}
2nd order     & Type I   & & 0.7688 & 0.6623 & 0.5753 & & 0.8405 & 0.7533 & 0.6662 & & 0.0109 & 0.0033 & 0.0011 & & 0.0181 & 0.0070 & 0.0029 \\
neighbourhood & Type II  & & 0.7910 & 0.6994 & 0.6210 & & 0.8546 & 0.7793 & 0.7080 & & 0.0092 & 0.0027 & 0.0008 & & 0.0172 & 0.0070 & 0.0031 \\
& Type III & & 0.7717 & 0.6671 & 0.5824 & & 0.8449 & 0.7501 & 0.6636 & & 0.0027 & 0.0004 & 0.0001 & & 0.0054 & 0.0010 & 0.0002 \\
& Type IV  & & {\bf 0.8245} & {\bf 0.7445} & {\bf 0.6775} & & {\bf 0.8812} & {\bf 0.8124} & {\bf 0.7462} & & 0.0033 & 0.0006 & 0.0001 & & 0.0059 & 0.0012 & 0.0003  \\
\noalign{\smallskip}\hline
\end{tabular}}
\end{center}
\end{table}

\end{landscape}

\subsection{Numerical simulation}
\label{sec:SimulationStudy2}
In this section we want to evaluate the computational gain offered by the scalable modelling approach against the Global model as the number of small areas increases. Specifically, we simulate a regular grid map with number of areas equal to $n=256,1024 \mbox{ and } 4096$, while the number of time points have been fixed to $T=25$ to imitate the real data analysis. For each template, spatially structured (CAR), temporally structured (RW1) and completely structured spatio-temporal (Type IV) random effects are generated from the corresponding structure matrices to define a log-risk surface (see Equation~\eqref{eq:Global_model}). Finally, we simulate counts for each areal-unit from a Poisson distribution as described in the previous section. To fit the scalable models, a $4 \times 4$ regular grid is used to define the partition of the spatial domain, so that a total of $D=16$ local spatio-temporal models are fitted. These models are distributed over 4 machines with 4 models running in parallel for each machine using the \texttt{bigDM} package.

\begin{figure}[!ht]
\begin{center}
\vspace{0.5cm}
\includegraphics{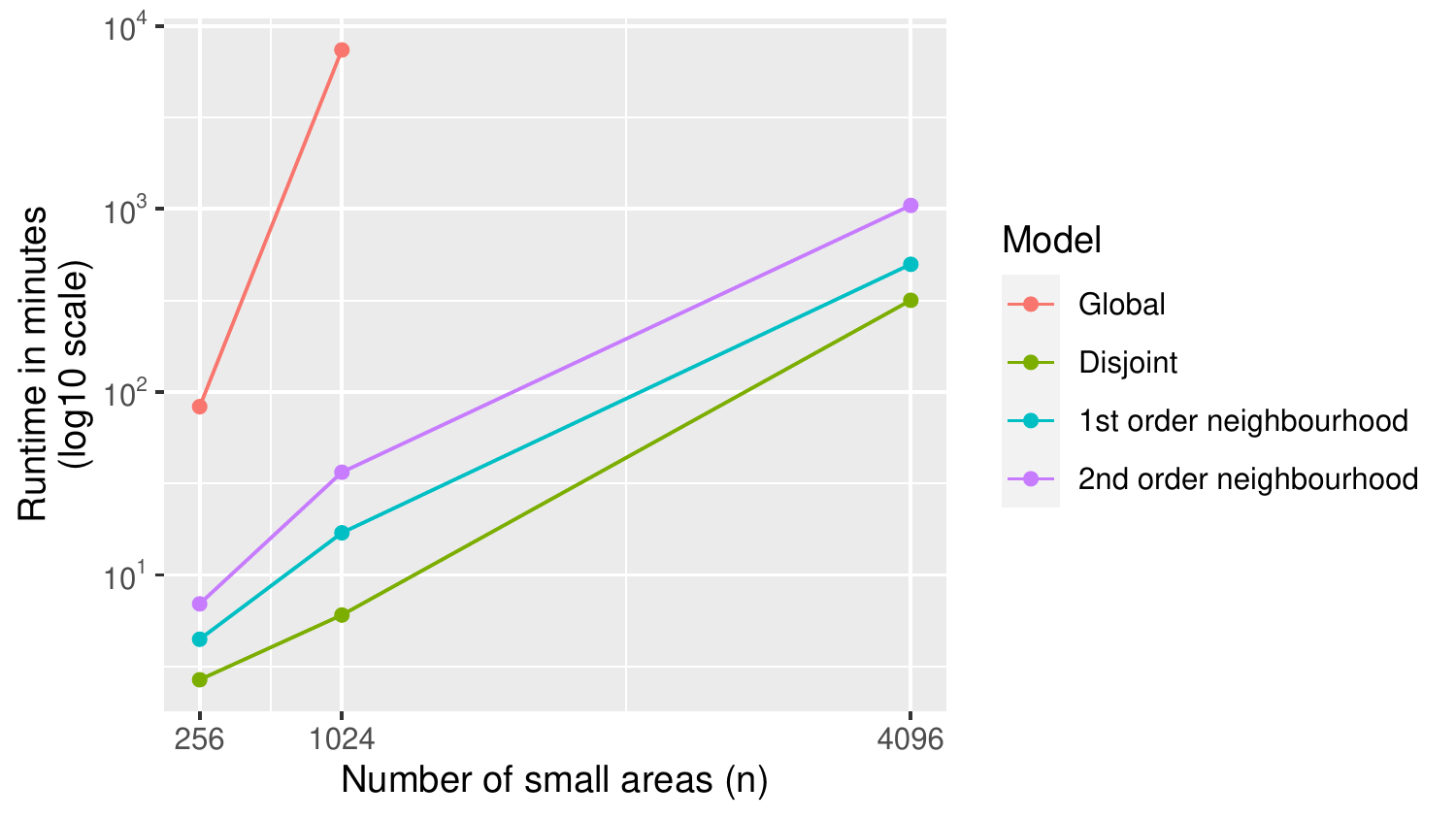}\hfill
\vspace{-0.5cm}
\end{center}
\caption{Numerical simulation: computational time (in log10 scale) vs number of small areas for the Global model and our scalable modelling proposals. \label{fig:Numerical_simulation}}
\end{figure}

In \autoref{fig:Numerical_simulation}, we show the total runtime of the different models when varying the number of spatial areas. As we increase the dimension of the spatial domain, the Global model quickly becomes computationally prohibitive. For $n=256$ areas the total running time is about 80 minutes, while for $n=1024$ areas the computational time exceeds 120 hours. For larger area sizes considered here, computation fails due to very high RAM memory usage. On the other hand, we can fit the Disjoint and k-order neighbourhood models for $n=256$ areas in total running times between 2-6 minutes, for $n=1024$ areas in times between 6-36 minutes, and for $n=4096$ areas in running times between 5-16 hours.

\section{Data analysis: lung cancer mortality in Spain}
\label{sec:Data_analysis}
We illustrate and compare all the approaches described in this paper by modelling the spatio-temporal evolution of male lung cancer mortality data in the $n=7907$ municipalities of continental Spain (excluding Balearic and Canary Islands and the autonomous cities of Ceuta and Melilla) during the period 1991-2015. According to recent studies \citep{ferlay2018cancer}, lung cancer was the leading cause of cancer deaths among the male population and the second cause among the female population in Europe in 2018, representing 24.8\% and 14.2\% of all cancer deaths, respectively. It also was the leading cause of cancer related deaths in Spain for both sexes in 2017, representing the 19.5\% of cancer mortality \citep{de2020cifras}. One of the main causes is that Spain is a country with a traditionally high tobacco consumption, with a smoking rate of over 32\% of the population at the beginning of the century \citep{remon2021lung}.

A total of 378,720 lung cancer deaths (corresponding to International Classification of Diseases-10 codes C33-C34) were registered for male population in the municipalities of continental Spain during the period 1991-2015 (that account for around 26\% of all malignant tumours for the target population during our study period), where the number of observed deaths per areal-time unit varies from 0 to 1247 (with a mean value of 1.9). The number of expected cases were computed using the indirect (internal) standardization method with 5-years age groups as standardization variable. The number of expected deaths per areal-time unit varies from 0 to 1332 (with a mean value of 1.9). A brief summary of the number of observed deaths, expected deaths and standardized mortality ratios for the provincial capital municipalities during the whole period is displayed in \autoref{tab:LungCancer_CrudeRates}.

\begin{table}[!b]
\caption{Lung cancer observed deaths, expected deaths and standardized mortality ratios (SMR) for the provincial capital municipalities during the period 1991-2015. \label{tab:LungCancer_CrudeRates}}
\vspace{-0.2cm}
\renewcommand{\arraystretch}{1.2}
\begin{center}
\resizebox{\textwidth}{!}{
\begin{tabular}{lrrr|lrrr|lrrr}
\hline\noalign{\smallskip}
Municipality & Obs. & Exp. & SMR & Municipality & Obs. & Exp. & SMR & Municipality & Obs. & Exp. & SMR \\
\hline
\'Avila         &  381 &   467.2 & 0.815 & Salamanca       & 1568 &  1621.7 & 0.967 & Alicante        & 2899 &  2634.2 & 1.101 \\
Segovia         &  443 &   533.4 & 0.831 & Palencia        &  760 &   785.0 & 0.968 & Barcelona       &18161 & 16434.4 & 1.105 \\
Burgos          & 1335 &  1572.0 & 0.849 & Girona          &  644 &   663.5 & 0.971 & Almer\'ia       & 1491 &  1324.4 & 1.126 \\
Vitoria         & 1795 &  2053.5 & 0.874 & Lugo            &  886 &   903.0 & 0.981 & A Coru\~na      & 2693 &  2385.2 & 1.129 \\
Logro\~no       & 1098 &  1231.8 & 0.891 & San Sebasti\'an & 1750 &  1781.0 & 0.983 & Zaragoza        & 6875 &  6083.2 & 1.130 \\
Guadalajara     &  576 &   644.9 & 0.893 & Madrid          &28750 & 29048.5 & 0.990 & Santander       & 2049 &  1786.6 & 1.147 \\
Cuenca          &  405 &   452.6 & 0.895 & Valladolid      & 3034 &  3060.3 & 0.991 & Sevilla         & 6605 &  5667.2 & 1.165 \\
Ja\'en          &  784 &   876.3 & 0.895 & Tarragona       & 1063 &  1066.8 & 0.996 & Valencia        & 8261 &  7046.9 & 1.172 \\
Soria           &  322 &   357.2 & 0.901 & Pamplona        & 1837 &  1804.1 & 1.018 & Ciudad Real     &  610 &   513.7 & 1.187 \\
Albacete        & 1103 &  1208.5 & 0.913 & Murcia          & 3048 &  2982.6 & 1.022 & Oviedo          & 2450 &  2026.9 & 1.209 \\
Ourense         & 1020 &  1104.3 & 0.924 & Pontevedra      &  669 &   646.7 & 1.034 & M\'alaga        & 5122 &  4213.2 & 1.216 \\
Zamora          &  618 &   668.4 & 0.925 & Toledo          &  613 &   579.6 & 1.058 & C\'aceres       &  839 &   653.6 & 1.284 \\
Granada         & 1940 &  2094.7 & 0.926 & L\'erida        & 1164 &  1094.9 & 1.063 & Huelva          & 1447 &  1089.6 & 1.328 \\
Teruel          &  298 &   318.0 & 0.937 & Cordoba         & 2767 &  2591.8 & 1.068 & Badajoz         & 1445 &  1055.5 & 1.369 \\
Huesca          &  449 &   472.7 & 0.950 & Castell\'on     & 1427 &  1320.7 & 1.080 & C\'adiz         & 1637 &  1164.4 & 1.406 \\
Le{\'o}n        & 1406 &  1455.5 & 0.966 & Bilbao          & 4053 &  3702.6 & 1.095 \\
\noalign{\smallskip}\hline
\end{tabular}}
\end{center}
\end{table}

For the Disjoint and 1st/2nd-order neighbourhood models presented in this section, we distribute the models over 7 machines with Intel Xeon E5-2620 v4 processors and 256GB RAM on each machine (CentOS Linux release 7.3.1611 operative system), using the simplified Laplace approximation strategy in R-INLA (stable version INLA 22.05.07) of R-4.1.3 and simultaneously running 5 models in parallel on each machine using the \texttt{bigDM} package. Again, it was not possible to fit Type II and Type IV interaction Global models using a single machine with the described characteristics. Results in terms of model selection criteria and computational time are shown in \autoref{tab:LungCancer_Results}. For the scalable models only the results regarding the \texttt{"original"} merging strategy are shown, since the simulation study shows that this procedure outperforms the \texttt{"mixture"} strategy in terms of risk estimation accuracy and high/low risk area detection.

\begin{figure}[!ht]
\begin{center}
\includegraphics[width=1.1\textwidth]{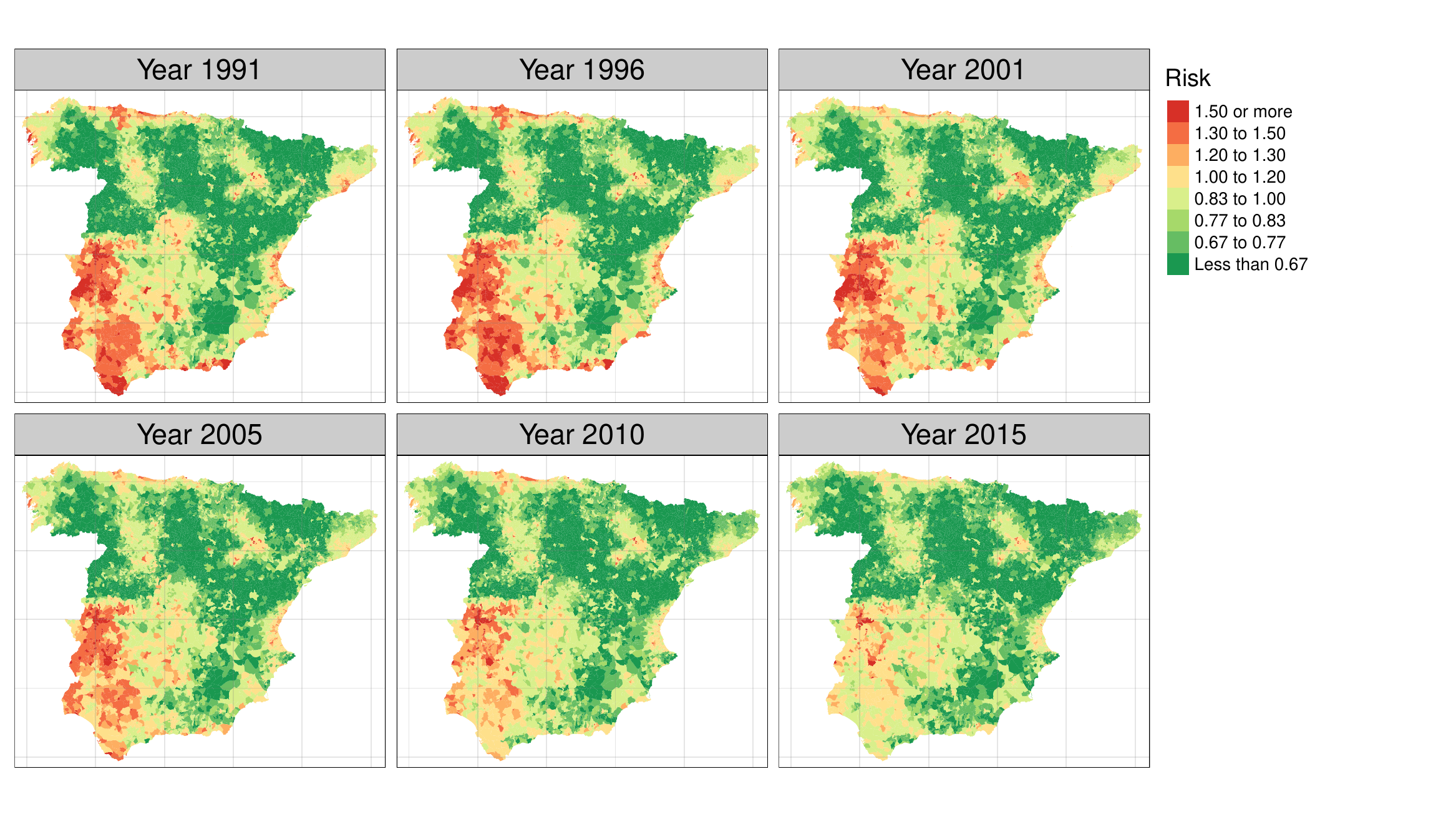}\hfill
\includegraphics[width=1.1\textwidth]{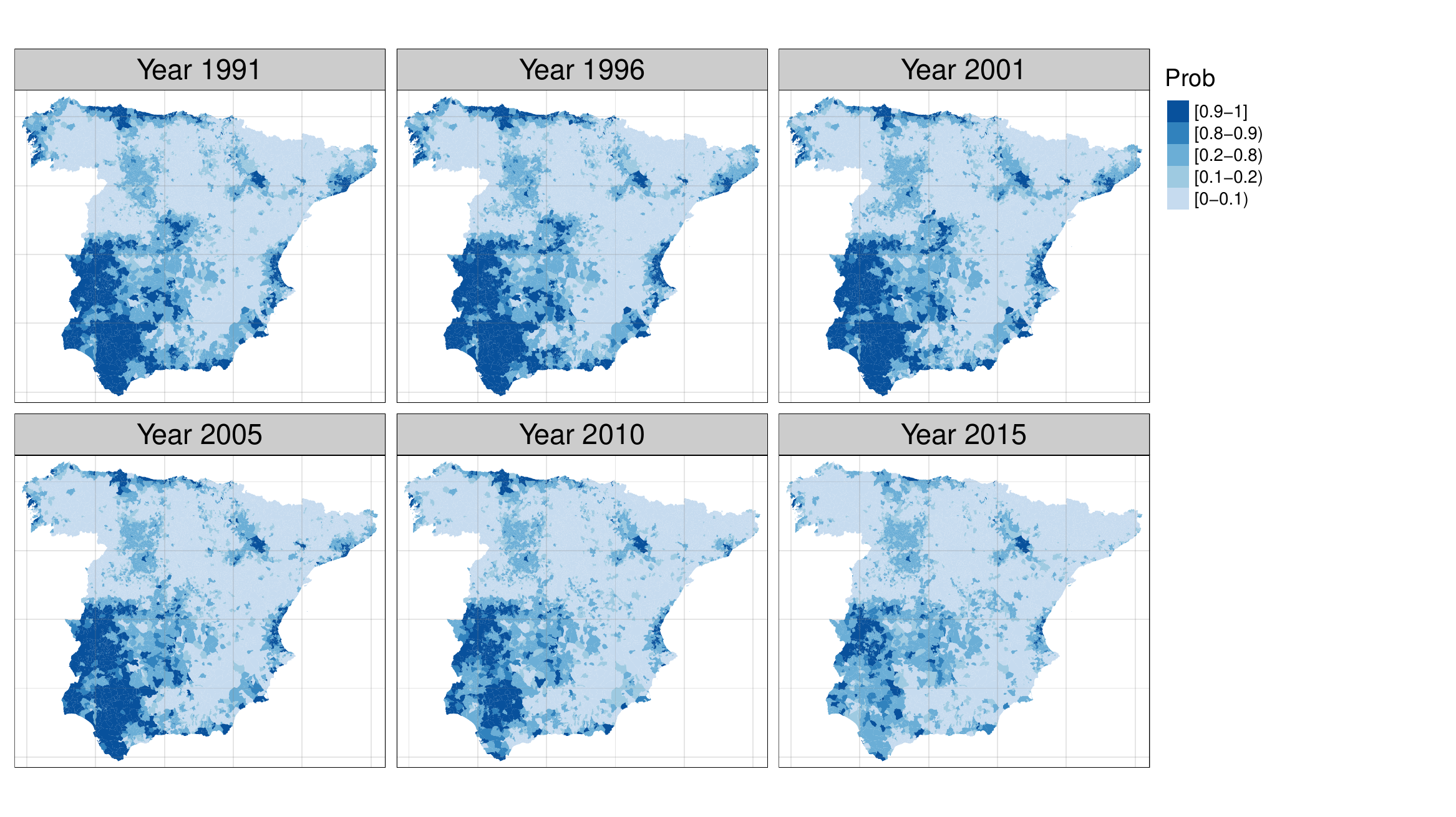}\hfill
\vspace{-1cm}
\end{center}
\caption{Maps of posterior median estimates of relative risks $r_{it}$ (top) and posterior exceedence probabilities $P(r_{it}>1 | {\bf O} )$ (bottom) for the 1st-order neighbourhood model considering a BYM2 conditional autoregressive prior for space, RW1 prior for time and Type IV interaction for the spatio-temporal effect. \label{fig:LungCancer_EstimatedRisks}}
\end{figure}

It can be seen that both DIC and WAIC model selection criteria support Type IV and Type II interaction effects, which precisely are those that cannot be fitted with the Global model. Besides the computational gain, the scalable model proposals are better supported by fit measures. In particular, 1st-order neighbourhood models shows slightly better performance.
Maps with the posterior median estimates of relative risks and posterior exceedence probabilities $P(r_{it}>1 | {\bf O})$ obtained with the 1st-order neighbourhood model considering a Type IV interaction for the spatio-temporal random effect are plotted at \autoref{fig:LungCancer_EstimatedRisks}. The estimated risk surfaces are consistent with those described by \cite{lopez2014time}, where the geographical pattern of lung cancer mortality data in Spain at municipality level was analyzed using spatial models.

\begin{table}[!t]
\caption{Model selection criteria and computational time (in minutes) for models fitted using the simplified Laplace approximation strategy of INLA and the \texttt{"original"} merging strategy. \label{tab:LungCancer_Results}}
\renewcommand{\arraystretch}{1}
\begin{center}
\resizebox{\textwidth}{!}{
\begin{tabular}{llrrrrrrr}
\hline\noalign{\smallskip}
Model & Interaction & $\overline{{D(\theta)}}$ & $p_{D}$ & DIC & WAIC & T.run & T.merge & T.total \\
\noalign{\smallskip}\hline\noalign{\smallskip}
Global        & Type I   & 144680 & 2984 & 147664 & 147696 & 663 & - & 663 \\
              & Type II  & - & - & - & - & - & - & - \\
              & Type III & 144467 & 2968 & 147435 & 147458 & 3846 & - & 3846 \\
              & Type IV  & - & - & - & - & - & - & - \\
\noalign{\smallskip}\hline\noalign{\smallskip}
Disjoint      & Type I   & 143154 & 3999 & 147154 & 147161 & 10  & 6 & 16  \\
              & Type II  & 143175 & 3801 & 146976 & 147045 & 218 & 6 & 224 \\
              & Type III & 143101 & 4015 & 147116 & 147161 & 22  & 6 & 27  \\
              & Type IV  & 143131 & 3753 & 146884 & 146965 & 259 & 6 & 264 \\
\noalign{\smallskip}\hline\noalign{\smallskip}
1st order     & Type I   & 143269 & 3824 & 147094 & 147112 & 14   & 8  &  22 \\
neighbourhood & Type II  & 143255 & 3671 & 146926 & 146997 & 548  & 8  & 557 \\
              & Type III & 143497 & 3562 & 147058 & 147123 & 34   & 8  &  42 \\
              & Type IV  & 143370 & 3458 & 146828 & 146910 & 636  & 8  & 644 \\
\noalign{\smallskip}\hline\noalign{\smallskip}
2nd order     & Type I   & 143500 & 3603 & 147103 & 147138 & 19   & 10 &   28 \\
neighbourhood & Type II  & 143366 & 3566 & 146932 & 147009 & 1740 & 10 & 1750 \\
              & Type III & 143731 & 3331 & 147062 & 147130 & 59   & 10 &   68 \\
              & Type IV  & 143523 & 3307 & 146830 & 146912 & 1879 & 10 & 1889 \\
\noalign{\smallskip}\hline\\[-2.ex]
\multicolumn{9}{l}{$\overline{D({\theta})}$: mean deviance, $p_{D}$: effective number of parameters}\\
\multicolumn{9}{l}{DIC: deviance information criterion, WAIC:  Watanabe-Akaike information criterion}\\
\multicolumn{9}{l}{T.run: running time, T.merge: merging time, T.total: running + merging time}
\end{tabular}}
\end{center}
\end{table}

\section{Conclusions}
\label{sec:Conclusions}

The use of spatial and spatio-temporal hierarchical models for regional data are crucial in areas such as cancer epidemiology, since they allow to obtain reliably incidence or mortality risk estimates of cancer in small areas, avoiding the huge variability of classical risk estimation measures such as the standardized mortality ratios or the crude rates. Research in this area has been very fruitful in recent decades and numerous statistical models have been proposed to study the geographic distribution of cancer and its evolution in time, as well as the underlying spatio-temporal patterns. However, the scalability of these models, i.e., their use when the number of areas increases significantly, has not been studied in depth yet. For that reason, the pragmatic, simple, and useful methodology proposed in this paper aims to provide alternative modelling approaches to disease mapping models commonly used when analysing high-dimensional spatio-temporal data.

Despite the enormous expansion of modern computers and the development of new software and estimation techniques to make fully Bayesian inference, dealing with massive data is still computationally challenging. Our proposal is based on the
on the idea of ``divide-and-conquer'' so that local spatio-temporal models can be simultaneously fitted. Adapting this idea to the context of disease mapping is very appropriate when the number of small areas is very large for three main reasons: (1) it is a natural and ``simple'' strategy, (2) the larger the spatial domain is, the less likely it is that the data are stationary across the whole map, and (3) it provides a scalable modelling scheme that substantially reduces the RAM/CPU memory usage and computational time.

Our simulation study indicates that the proposed methodology provides reliable risk estimates with a substantial reduction in computational time. 
Futhermore, we observe that our model proposals perform better in detecting high/low risk areas, by obtaining higher true positive and true negative rates than when considering the usual spatio-temporal CAR models, avoiding false alarms.
Regarding the merging strategy of the areas belonging to different subdomains, we compare the use of mixture distributions to combine the posterior marginal density functions against using the posterior marginal estimate of the areal-unit corresponding to the original subdomain. Our simulation study shows that the latter strategy (denoted as \texttt{"original"} merging strategy) reduces computational time while providing better results in terms of risk estimation accuracy and true positive/negative rates. On the other hand, in some cases it may not be sufficient to use first-order neighbours to avoid the boundary effect caused by the division of the whole study region into smaller subdomains.
%
We have additionally analyzed the advantages of our scalable model proposal in terms of computational complexity as the number of small areas increases. Our numerical simulation study shows a substantial reduction in computational time in comparison with the Global models.
Finally, lung cancer mortality data in the municipalities of Spain during the period 1991-2015 have been analyzed to illustrate the new model proposals, using the administrative division of continental Spain in 47 provinces to define the partition of the spatial domain. Doing so, we are able to fit a CAR model that accounts for both spatial and temporal dependence by including completely structured space-time interaction random effects (commonly denoted as Type IV interaction), which was computationally unfeasible to fit when considering non-scalable models.

The methods and algorithms proposed in this work are implemented in the open-source \texttt{R} package \texttt{bigDM} (\url{https://cran.r-project.org/web/packages/bigDM/index.html}). This package allows the user to adapt the modelling scheme to their own processing architecture by performing both parallel and/or distributed computation strategies to speed up computations by using the \texttt{future} package.
Model fitting and inference is carried out using INLA methodology through the \texttt{R-INLA}, as it is now a well-known Bayesian approximation technique, computationally efficient and easy for practitioners to handle. Very recently, promising research in a hybrid approximate method that uses the Laplace method with a low-rank Variational Bayes correction to the posterior mean has been released \citep{vanNiekerk2021,vanNiekerk2022}. This new approximation technique has been shown to provide accurate results with a superior computational efficiency and scalability than the classic integrated nested Laplace approximations. Currently, it is implemented in \texttt{R-INLA} as an experimental mode, but as stated by the developers of this technique, it will presumably be enabled by default in the near future. When the moment comes, we plan to adapt our \texttt{bigDM} package to be compatible with this new avenue for Bayesian inference with INLA.

Finally, we are currently working on extending our Bayesian modelling proposal to ecological regression models that takes into account the spatial and/or spatio-temporal confounding issues between fixed and random effects \citep{adin2021alleviating}, as well as to high-dimensional multivariate disease mapping models in which several diseases are jointly analyzed \citep{Vicente2021}.

%

\section*{Acknowledgements}
\addcontentsline{toc}{section}{Acknowledgements}
This research has been supported by the project PID2020-113125RB-I00/MCIN/AEI/10.13039/501100011033. It has also been partially funded by the Public University of Navarra (project PJUPNA20001). We would like to thank Andrea Riebler and James Hodges for their useful comments that have contributed to improve this paper.


\bibliographystyle{apalike}
\bibliography{biblio}   

%
%

\newpage

\setcounter{section}{0} 
\renewcommand{\thesection}{\Alph{section}}

\setcounter{figure}{0}
\renewcommand\thefigure{\thesection.\arabic{figure}}

\setcounter{table}{0}
\renewcommand\thetable{\thesection.\arabic{table}}

\section{Appendix}
\label{Appendix}

\begin{table}[!ht]
\caption{Simulation study: average values of mean absolute relative bias (MARB), mean relative root mean square error (MRRMSE) and Interval Score (IS) for the 1st/2nd-order neighbourhood models {\bf computed only for the border areas}. \label{tab:Appendix1}}
\begin{center}
\resizebox{\textwidth}{!}{
\begin{tabular}{llrrrrrrr}
\hline\noalign{\smallskip}
& & \multicolumn{3}{c}{merge.strategy=``mixture"} & & \multicolumn{3}{c}{merge.strategy=``original"}\\
\cline{3-5} \cline{7-9}\\
Model & Interaction & MARB & MRRMSE & IS & & MARB & MRRMSE & IS \\
\noalign{\smallskip}\hline\noalign{\smallskip}
1st order     & Type I   & 0.0304 & 0.0430 & 0.2733 & & 0.0375 & 0.0471 & 0.2656 \\
neighbourhood & Type II  & 0.0264 & 0.0410 & 0.2394 & & 0.0318 & 0.0446 & 0.2305 \\
              & Type III & 0.0178 & 0.0369 & 0.2759 & & 0.0211 & 0.0389 & 0.2405 \\
              & Type IV  & 0.0141 & 0.0322 & 0.2140 & & 0.0154 & 0.0326 & 0.1906 \\
\noalign{\smallskip}\hline\noalign{\smallskip}
2nd order     & Type I   & 0.0323 & 0.0438 & 0.2734 & & 0.0359 & 0.0454 & 0.2576 \\
neighbourhood & Type II  & 0.0276 & 0.0422 & 0.2432 & & 0.0298 & 0.0435 & 0.2287 \\
              & Type III & 0.0181 & 0.0369 & 0.2715 & & 0.0181 & 0.0356 & 0.2409 \\
              & Type IV  & 0.0141 & 0.0320 & 0.2083 & & 0.0138 & 0.0308 & 0.1900 \\
\noalign{\smallskip}\hline
\end{tabular}}
\end{center}
\end{table}

\begin{table}[!ht]
\caption{Simulation study: average values of true and false positive rates for the reference threshold values of $p_0=0.8, 0.9 \mbox{ and } 0.95$, based on posterior exceedence probabilities $P(r_{it}>1|{\bf O})$. Results for the 1st/2nd-order neighbourhood models {\bf computed only for the border areas}. \label{tab:Appendix2A}}
\vspace{-0.5cm}
\begin{center}
\resizebox{\textwidth}{!}{
\begin{tabular}{llrrrrrrr}
\hline\noalign{\smallskip}
& & \multicolumn{6}{c}{\bf True Positive Rate} \\
\hline\noalign{\smallskip}
& & \multicolumn{3}{c}{merge.strategy=``mixture"} & & \multicolumn{3}{c}{merge.strategy=``original"}\\
\cline{3-5} \cline{7-9}\\
& & $p_0=0.8$ & $p_0=0.9$ & $p_0=0.95$ & & $p_0=0.8$ & $p_0=0.9$ & $p_0=0.95$ \\
\noalign{\smallskip}\hline\noalign{\smallskip}
1st order     & Type I   & 0.7598 & 0.6490 & 0.5494 & & 0.7789 & 0.6848 & 0.6024 \\
neighbourhood & Type II  & 0.7864 & 0.6939 & 0.6104 & & 0.8051 & 0.7255 & 0.6549 \\
              & Type III & 0.7609 & 0.6513 & 0.5545 & & 0.7849 & 0.6901 & 0.6106 \\
              & Type IV  & 0.8122 & 0.7302 & 0.6599 & & 0.8313 & 0.7591 & 0.6972 \\
\noalign{\smallskip}\hline\noalign{\smallskip}
2nd order     & Type I   & 0.7601 & 0.6459 & 0.5442 & & 0.7745 & 0.6757 & 0.5916 \\
neighbourhood & Type II  & 0.7840 & 0.6889 & 0.6023 & & 0.7973 & 0.7121 & 0.6379 \\
              & Type III & 0.7657 & 0.6583 & 0.5636 & & 0.7836 & 0.6875 & 0.6071 \\
              & Type IV  & 0.8185 & 0.7384 & 0.6703 & & 0.8312 & 0.7572 & 0.6950 \\
\noalign{\smallskip}\hline\\[-1.5ex]
& & \multicolumn{6}{c}{\bf False Positive Rate} \\
\hline\noalign{\smallskip}
& & \multicolumn{3}{c}{merge.strategy=``mixture"} & & \multicolumn{3}{c}{merge.strategy=``original"}\\
\cline{3-5} \cline{7-9}\\
& & $p_0=0.8$ & $p_0=0.9$ & $p_0=0.95$ & & $p_0=0.8$ & $p_0=0.9$ & $p_0=0.95$ \\
\noalign{\smallskip}\hline\noalign{\smallskip}
1st order     & Type I   & 0.0054 & 0.0011 & 0.0002 & & 0.0142 & 0.0052 & 0.0021 \\
neighbourhood & Type II  & 0.0053 & 0.0012 & 0.0003 & & 0.0119 & 0.0043 & 0.0015 \\
              & Type III & 0.0021 & 0.0003 & 0.0000 & & 0.0045 & 0.0009 & 0.0002 \\
              & Type IV  & 0.0027 & 0.0004 & 0.0001 & & 0.0044 & 0.0009 & 0.0002 \\
\noalign{\smallskip}\hline\noalign{\smallskip}
2nd order     & Type I   & 0.0062 & 0.0013 & 0.0003 & & 0.0125 & 0.0042 & 0.0016 \\
neighbourhood & Type II  & 0.0060 & 0.0015 & 0.0004 & & 0.0100 & 0.0031 & 0.0010 \\
              & Type III & 0.0024 & 0.0004 & 0.0001 & & 0.0029 & 0.0005 & 0.0001 \\
              & Type IV  & 0.0029 & 0.0005 & 0.0001 & & 0.0031 & 0.0006 & 0.0001 \\
\noalign{\smallskip}\hline
\end{tabular}}
\end{center}
\end{table}

\begin{table}[!ht]
\caption{Simulation study: average values of true and false negative rates for the reference threshold values of $p_0=0.8, 0.9 \mbox{ and } 0.95$, based on posterior exceedence probabilities $P(r_{it}<1|{\bf O})$. Results for the 1st/2nd-order neighbourhood models {\bf computed only for the border areas}. \label{tab:Appendix2B}}
\vspace{-0.5cm}
\begin{center}
\resizebox{\textwidth}{!}{
\begin{tabular}{llrrrrrrr}
\hline\noalign{\smallskip}
& & \multicolumn{6}{c}{\bf True Negative Rate} \\
\hline\noalign{\smallskip}
& & \multicolumn{3}{c}{merge.strategy=``mixture"} & & \multicolumn{3}{c}{merge.strategy=``original"}\\
\cline{3-5} \cline{7-9}\\
& & $p_0=0.8$ & $p_0=0.9$ & $p_0=0.95$ & & $p_0=0.8$ & $p_0=0.9$ & $p_0=0.95$ \\
\noalign{\smallskip}\hline\noalign{\smallskip}
1st order     & Type I   & 0.8063 & 0.6923 & 0.5954 & & 0.8291 & 0.7431 & 0.6580 \\
neighbourhood & Type II  & 0.8337 & 0.7359 & 0.6523 & & 0.8499 & 0.7768 & 0.7084 \\
              & Type III & 0.8182 & 0.6983 & 0.5943 & & 0.8367 & 0.7427 & 0.6576 \\
              & Type IV  & 0.8618 & 0.7774 & 0.6984 & & 0.8753 & 0.8060 & 0.7405 \\
\noalign{\smallskip}\hline\noalign{\smallskip}
2nd order     & Type I   & 0.8139 & 0.7008 & 0.6030 & & 0.8331 & 0.7470 & 0.6622 \\
neighbourhood & Type II  & 0.8388 & 0.7424 & 0.6571 & & 0.8507 & 0.7751 & 0.7043 \\
              & Type III & 0.8293 & 0.7148 & 0.6109 & & 0.8445 & 0.7494 & 0.6622 \\
              & Type IV  & 0.8725 & 0.7938 & 0.7178 & & 0.8824 & 0.8137 & 0.7476 \\
\noalign{\smallskip}\hline\\[-1.5ex]
& & \multicolumn{6}{c}{\bf False Negative Rate} \\
\hline\noalign{\smallskip}
& & \multicolumn{3}{c}{merge.strategy=``mixture"} & & \multicolumn{3}{c}{merge.strategy=``original"}\\
\cline{3-5} \cline{7-9}\\
& & $p_0=0.8$ & $p_0=0.9$ & $p_0=0.95$ & & $p_0=0.8$ & $p_0=0.9$ & $p_0=0.95$ \\
\noalign{\smallskip}\hline\noalign{\smallskip}
1st order     & Type I   & 0.0111 & 0.0029 & 0.0007 & & 0.0225 & 0.0091 & 0.0036 \\
neighbourhood & Type II  & 0.0108 & 0.0029 & 0.0007 & & 0.0209 & 0.0091 & 0.0042 \\
              & Type III & 0.0040 & 0.0005 & 0.0001 & & 0.0084 & 0.0019 & 0.0004 \\
              & Type IV  & 0.0043 & 0.0007 & 0.0001 & & 0.0073 & 0.0016 & 0.0004 \\
\noalign{\smallskip}\hline\noalign{\smallskip}
2nd order     & Type I   & 0.0124 & 0.0030 & 0.0007 & & 0.0199 & 0.0071 & 0.0025 \\
neighbourhood & Type II  & 0.0123 & 0.0032 & 0.0009 & & 0.0176 & 0.0066 & 0.0027 \\
              & Type III & 0.0047 & 0.0007 & 0.0001 & & 0.0055 & 0.0010 & 0.0002 \\
              & Type IV  & 0.0052 & 0.0009 & 0.0002 & & 0.0055 & 0.0011 & 0.0002 \\
\noalign{\smallskip}\hline
\end{tabular}}
\end{center}
\end{table}

\end{document}